\documentclass{aa}
\usepackage[T1]{fontenc}
\usepackage{ae,aecompl,ulem}
\usepackage{xcolor,listings,multirow}
\usepackage{textcomp,pifont}

\usepackage{graphicx}   
\usepackage{color}

\usepackage{etoolbox}
\makeatletter
\patchcmd\@combinedblfloats{\box\@outputbox}{\unvbox\@outputbox}{}{%
  \errmessage{\noexpand\@combinedblfloats could not be patched}%
}%
\makeatother

\begin{document}

\title{Improving Hickson-like compact group finders \\
in redshift surveys: an implementation in the SDSS}
\author{Eugenia D\'iaz-Gim\'enez\inst{1,2}\thanks{eugeniadiazz@gmail.com}
\and Ariel Zandivarez\inst{1,2}
\and Antonela Taverna\inst{1,2,3}
}
\institute{ Instituto de Astronom\'ia Te\'orica y Experimental (IATE), CONICET-UNC.
\and Observatorio Astron\'omico de C\'ordoba (OAC), Universidad Nacional de C\'ordoba (UNC), Argentina.
\and Facultad de Matem\'atica, Astronom\'ia, F\'isica y Computaci\'on (FaMAF), UNC, C\'ordoba, Argentina.}
\date{\today}

\abstract
{}
{In this work we present an algorithm to identify compact groups (CGs) that closely follows Hickson's original aim and that improves the completeness of the samples of compact groups obtained from redshift surveys.}
{Instead of identifying CGs in projection first and then checking a velocity concordance criterion, we identify them directly in redshift space using Hickson-like criteria.
The methodology was tested on a mock lightcone of galaxies built from the outputs of a recent semi-analytic model of galaxy formation 
run on top of the Millennium Simulation I after scaling to represent the first-year Planck cosmology.}
{ The new algorithm identifies nearly twice as many CGs, no longer missing CGs that failed the isolation criterion because of velocity outliers lying in the isolation annulus.
The new CG sample picks up lower surface brightness groups, which are both looser and with fainter brightest galaxies, missed by the
classic method.

A new catalogue of compact groups from the Sloan Digital Sky Survey is the natural corollary of this study. The publicly available sample comprises $462$ observational groups with four or more galaxy members, of which $406$ clearly fulfil all the compact group requirements: compactness, isolation, and velocity concordance of all of their members. The remaining $56$ groups need further redshift information of potentially contaminating sources.
This constitutes the largest sample of groups that strictly satisfy
all the Hickson's criteria in a survey with available spectroscopic information.}
 {}
\titlerunning{Improving HCG finders in redshift surveys}
\maketitle

\section{Introduction}
Over the past 40 years, the astronomical community has devoted much time to the systematic search for compact groups (CGs). These peculiar small galaxy systems have proven to be a very powerful tool to understand galaxy interactions in dense environments, shaping our knowledge of galaxy evolution.

The most popular sample of CGs was identified by \cite{Hickson82}. This sample was a systematic search of CGs on plates of the Palomar Observatory Sky Survey that relied on a visual inspection. \citeauthor{Hickson82} established a set of rules that should be fulfilled for a galaxy system to be considered a compact group: compactness, isolation (with a relatively empty annulus surrounding the group), and population. These rules were defined to obtain small, isolated, and compact galaxy systems. In this way, \citeauthor{Hickson82} created the well-known Hickson Compact Group (HCG) sample that comprises 100 galaxy groups identified in projection on the sky. However, since only angular positions were used to identify the sample, their truly compact nature was questioned  (e.g. \citealt{Mamon86,Mamon90}). A few years later, the redshifts of their galaxy members were measured and a velocity filter was added to the criteria, resulting in a sample of 68 CGs with at least four concordant velocities \citep{Hickson92}.

Since then, several works have identified CGs in observational or simulated galaxy catalogues based on Hickson's recipes with or without including the velocity filtering depending on the available information (e.g. \citealt{Prandoni+94,Iovino+02,Lee+04,McConnachie+08,McConnachie+09,DiazGimenez&Mamon10,DiazGimenez+12,sohn+15}). Most of these works share the same philosophy for the algorithm construction, i.e. finding CGs that meet Hickson's criteria 
through an automatic procedure that resembles the original Hickson's visual inspection by first detecting candidate CGs on the sky, and then discarding obvious foreground/background galaxies along the line of sight from their discordant redshifts. 
However, for quasi-complete spectroscopic galaxy catalogues this procedure is not optimal because it may discard groups that appear not isolated on the sky, although the galaxies populating the isolation annulus turn out to all have discordant redshifts. 
Having chance projected galaxies in the cone is not a problem if  subsequent velocity filtering is performed. The big problem is that if such galaxies lie outside the CG, they will lead the algorithm to discard the group as non-isolated, whereas they are clear chance projections along the line of sight. 
The type of restrictions that arises from the procedure itself could easily bias subsequent studies related to CGs environment.

\begin{figure*}
\begin{center}
\includegraphics[width=9.1cm,trim=125 90 90 120,clip]{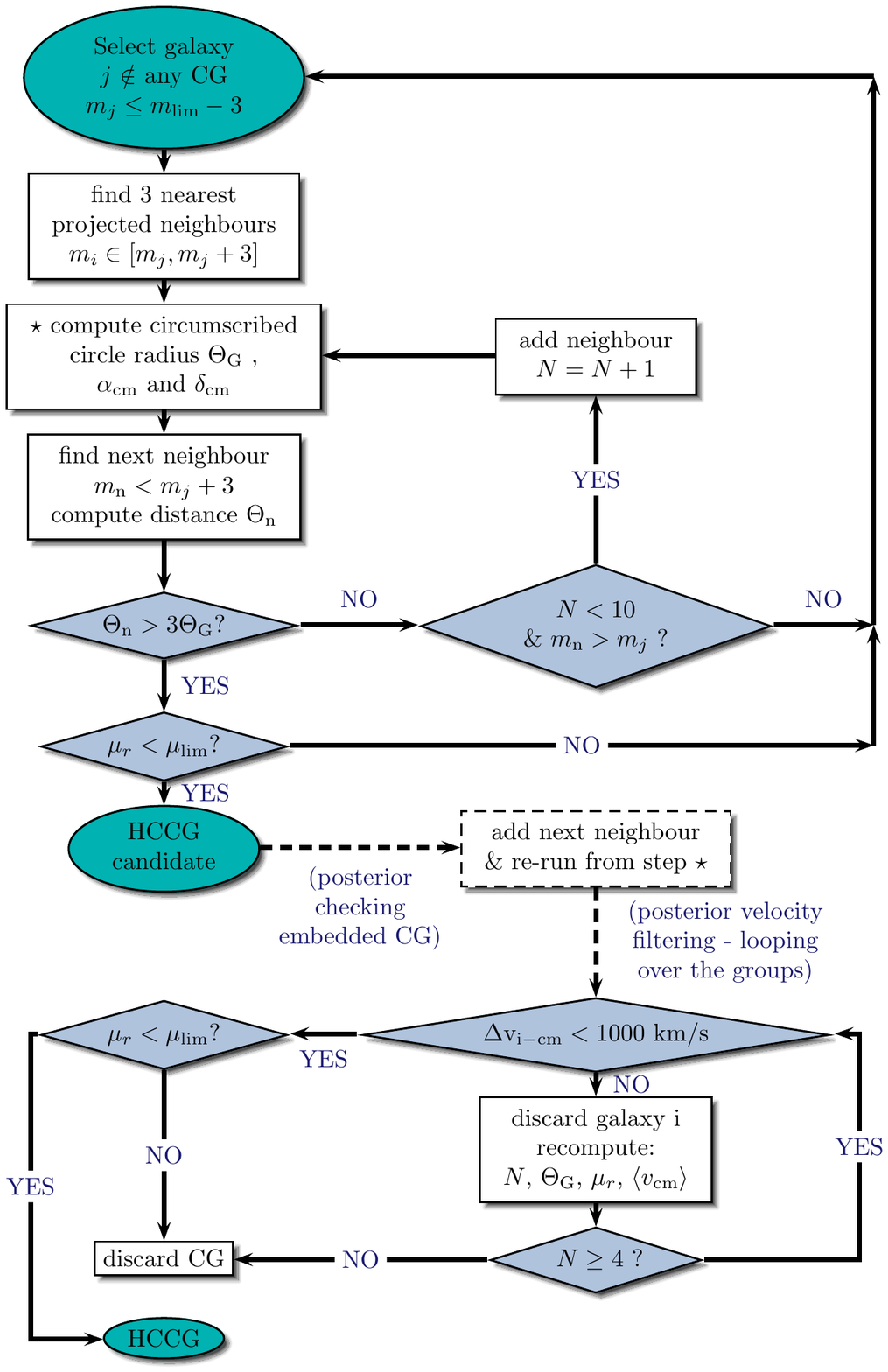}
\includegraphics[width=9.1cm,trim=130 155 120 120,clip]{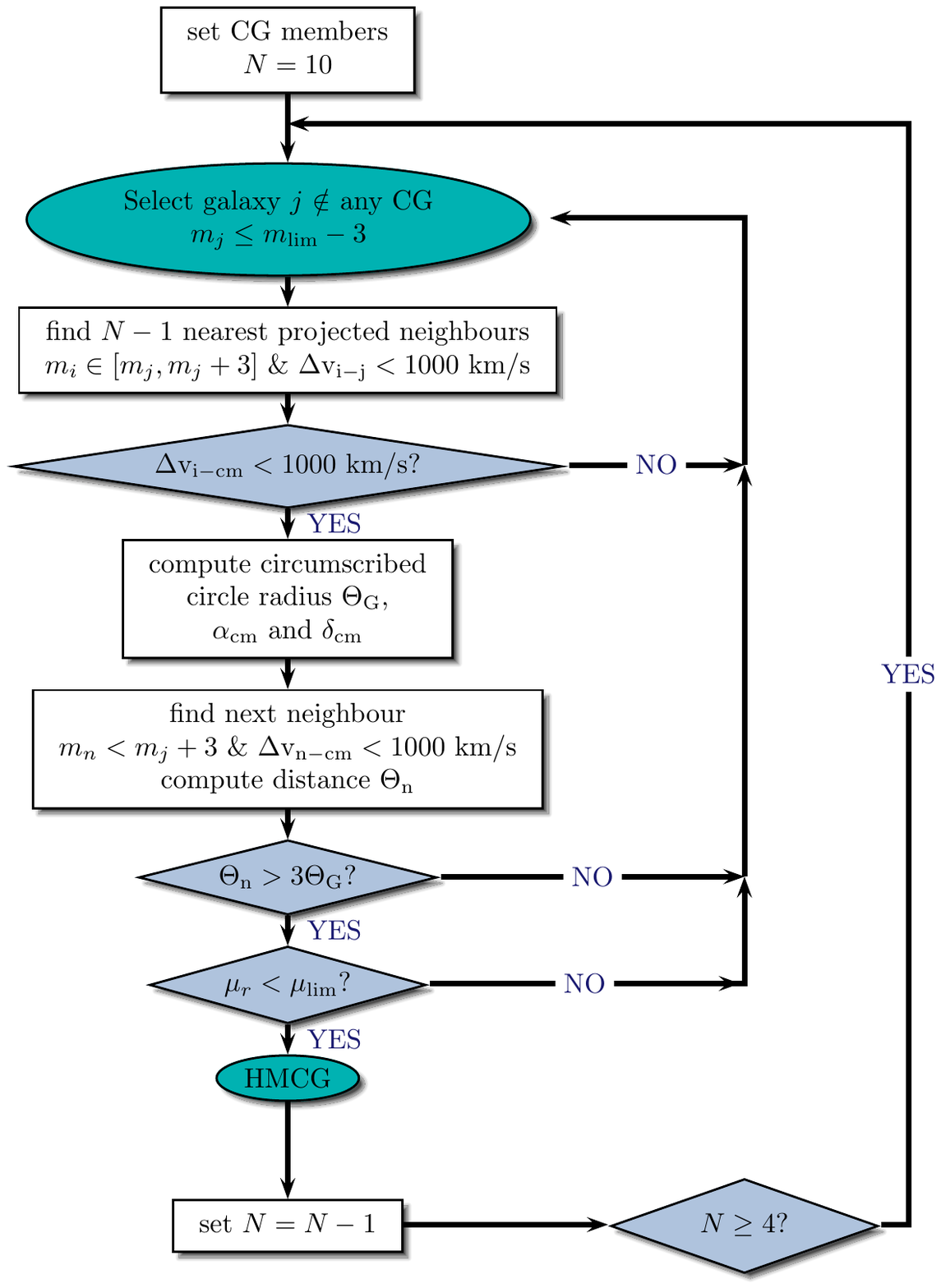}
\caption{\label{flux_diagram} Flow charts showing the automatic implementation of the Hickson's criteria to identify CGs. On the left the flow chart shows the \texttt{classic} algorithm, while on the right the \texttt{modified} algorithm is shown.
}
\end{center}
\end{figure*}

 In the past, some authors  already noticed this problem and suggested new ways to identify CGs in redshift surveys by changing the searching procedure and using percolation algorithms similar to the friends-of-friends (FoF) algorithm  (e.g. \citealt{Barton96,Focardi&Kelm02,Zandivarez03,sohn+16}). However, most of these attempts disregarded most or all of Hickson's definitions, and only kept a compactness criterion based on the physical size of the galaxy systems or the intergalactic separations.

The purpose of the present work is to present a \texttt{modified} algorithm that applies the Hickson's criteria directly in redshift space.
As a result, we present a new catalogue of CGs extracted from the Sloan Digital Sky Survey (SDSS) that satisfies Hickson's criteria. 
 
The layout of this work is as follows. In Section 2 we describe the construction of the mock galaxy lightcone used in this work for the testing process. This catalogue was constructed using synthetic galaxies obtained from a semi-analytical model (SAM) of galaxy formation \citep{Henriques+15}. In Section 3 we describe the well-known Hickson's recipe for identifying CGs \citep{Hickson82,Hickson92} and the previous and new algorithms for identifying CGs that fulfil these conditions. We apply these algorithms to the mock catalogue and perform the corresponding comparisons between the resulting CG catalogues. In Section 4 we apply both algorithms to an observational galaxy sample compiled from the SDSS by \cite{tempel17} and compare the resulting new CG sample with previous observational identifications. Finally, in Section 5 we summarise our results and present our conclusions. 

\section{Mock galaxy catalogue}
\label{sec:mock}
Using the publicly available outputs of the  SAM of \cite{Henriques+15} run on the N-body Millennium simulation \citep{Springel+05} re-scaled to the Planck cosmology\footnote{\url{http://www.mpa-garching.mpg.de/millennium/}}, 
we construct a mock lightcone galaxy catalogue following a  procedure similar to that described in \cite{jpas} (see  subsection 2.3 for details). 
The lightcone was built using galaxies extracted from different redshift outputs of the simulation to include the evolution of structures and galaxy properties with time. 

The lightcone is limited to an SDSS $r$-band observer-frame AB apparent magnitude of $r< 17.77$.  To compute the observer-frame magnitudes in the lightcones, it is necessary to k-decorrect the  rest-frame magnitudes provided by the SAM. An iterative process is used to compute the k-corrections,
and therefore the observer-frame magnitudes. A detailed description of this procedure is included in  Appendix~\ref{k-corr}. 

The final sample comprises $3 \, 139 \, 409$ galaxies within a solid angle of $4 \, \pi\,\rm sr$.

\section{Compact group finder algorithms}
\label{sec:sample}

Hickson-like CGs are expected to meet the following criteria: 
\begin{description} 
\item [{\bf Population}:] $$4 \le N \le 10;$$
\item [{\bf Compactness}:] $$\displaystyle \mu_r \le \mu_{\rm lim};$$
\item [{\bf Isolation}:] $$\displaystyle \Theta_n > 3 \,\Theta_G;$$
\item [{\bf Flux limit}:] $$\displaystyle r_{\rm b} \le r_{\rm lim} - 3;$$
\item [{\bf Velocity filtering}:]   $$ \Delta {\rm v}_{i,{\rm cm}}  =   \displaystyle c \, \frac{|z_i - \langle z_{\rm cm} \rangle|}{1+\langle z_{cm} \rangle }  \le  \displaystyle  1000 \, \rm km \, s^{-1} $$
\end{description}
Here, $N$ is the number of galaxies whose $r$-band magnitudes are within a three-magnitude range from the brightest galaxy;
$\mu_r$ is the mean $r$-band surface brightness averaged over 
the smallest circle that circumscribes the galaxy centres;
$\Theta_{\rm G}$ is the angular diameter of the smallest circumscribed circle; 
$\Theta_{\rm n}$ is the angular diameter of the largest
concentric circle that contains no other galaxies within the considered magnitude range or brighter;
$r_{\rm b}$ is the apparent magnitude of the brightest galaxy of the group; 
$r_{\rm lim}$ is the apparent magnitude limit of the parent catalogue, in our case $r_{\rm lim}=17.77$; 
$z_i$ is the spectroscopic redshift of each galaxy member;
and $\langle z_{\rm cm} \rangle$ is the median of the redshifts of the galaxy members. 
The flux limit criterion has to be included to ensure the membership and isolation of the groups since the population and isolation criteria are checked against galaxies within a three-magnitude range from the brightest \citep{DiazGimenez&Mamon10}. 
The value of $\mu_{\rm lim}$ (surface brightness limit) depends on the photometric band in which the selection is made. In our case, we work on the $r$ band; therefore, the adopted limit is $\mu_{\rm lim} =26.33$ [mag\,arcsec$^{-2}$] \citep{Taverna+16}.

Different algorithms can be applied in order to produce a sample of CGs that meets all the criteria described above. As we will show later, the order in which the criteria are applied affects the completeness of the resulting sample.

\subsection{Classic algorithm}

In several previous works \citep{DiazGimenez&Mamon10,DiazGimenez+12,Zandivarez+14,DiazGimenez+15,Taverna+16}, we   applied an algorithm to produce samples of Hickson-like compact groups. 
Following the original procedure, we  identified CGs in projection on the plane of sky that met the first four criteria mentioned above, and   we applied the velocity filtering only
in a second stage. 

The description of our algorithm can be found in \cite{DiazGimenez&Mamon10}. Hereafter, we  refer to this method as \texttt{HICKSON CLASSIC (HC)}. Its full implementation is shown in the flow chart in Fig.~\ref{flux_diagram} (left).  The algorithm starts looking for the smallest CGs; it starts with only four galaxies and it checks all the CG criteria. If the isolation is not fulfilled, it might eventually include new galaxy members or discard the CG. In a subsequent stage, it is checked for embedded compact groups: there could be CGs contained within 
a larger CG. We detected these cases, and adopted the largest ones. Finally, for these projected compact groups, a velocity filtering is applied through an iterative procedure that discards the farthest interloper and checks compactness, population, and velocity again.

When working with mock galaxy catalogues, it is also necessary to take into consideration that  galaxies in the mock catalogue are just point-sized particles, while observed galaxies are extended objects. Following \cite{DiazGimenez&Mamon10} we  therefore included the blending of galaxies in projection on the plane of the sky, which modifies the number of detectable galaxies and changes the population of CGs. To examine whether two galaxies are blended, we first computed the half-light radii in the $r$ band as a function of the stellar mass of each mock galaxy following the prescriptions of \cite{Lange+15} (see their eqs.~(2) and~(3)), using the bulge-to-total stellar mass ratio as a proxy for morphological type \citep{Bertone+07}. Finally, we considered that two galaxies are blended if their angular separation is smaller than the sum of their angular half-light radii. The number of members is recomputed considering the galaxies that have been blended, and groups whose population is less than four galaxies are discarded. The blending is checked twice, once when the projected compact groups are identified, and  again after applying the velocity filter. 

The application of the \texttt{classic} algorithm leads to a final sample that comprises $680$ mock Hickson classic compact groups (\texttt{mHCCG}s) in the lightcone described in Sect.~\ref{sec:mock} (see Table~\ref{tab:num}). 

\begin{figure*}
\centering
\includegraphics[width=0.8\textwidth,trim=0 110 50 0,clip]{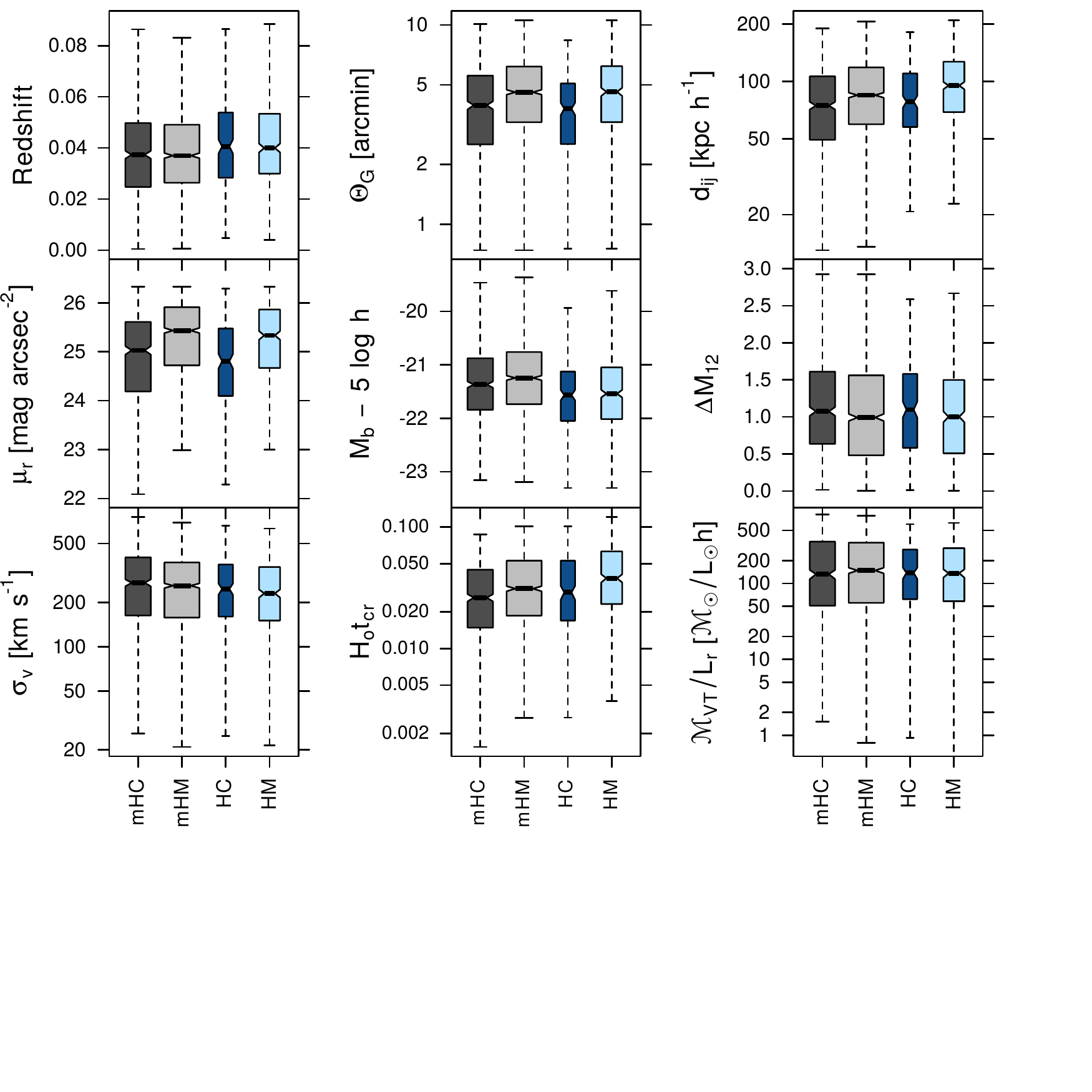}
\caption{Boxplot diagrams of CG properties for groups identified using the classic and the \texttt{modified} algorithm. The notches (waists) in the boxes indicate $\sim 95$ \%\ confidence intervals for the medians, while the widths are proportional to the square roots of the number of objects in each sample \citep{plotR}. The boxplot diagrams on the left side in each panel are the samples obtained using the \texttt{classic} and the \texttt{modified} algorithms in the mock catalogue (\texttt{mHC} and \texttt{mHM}), while those at the right side show the results for the samples obtained applying both algorithms on the SDSS catalogue (\texttt{HC} and \texttt{HM}).\label{fig:boxplot}}
\end{figure*}

\begin{table}
\caption{Number of CGs identified using the \texttt{classic} and \texttt{modified} algorithms. \label{tab:num}}
\begin{center}
\begin{tabular}{crr}
\hline
\hline
Samples & \multicolumn{2}{c}{Algorithm}\\
\cline{2-3}
& \texttt{Classic} & \texttt{Modified}\\
\hline
No. of mock CGs & 680 & 1287 \\
\% real CGs & 38 & 35 \\
\hline
 No. of SDSS CGs & 218 & 462\\
\hline
\hline
\end{tabular}
\end{center}
\end{table}

\subsection{Modified algorithm}
\label{modified}
Our implementation of the \texttt{classic} algorithm was inspired by the original path followed by Hickson at a time when redshift surveys were not available. Different authors were able to create new samples following the same automatic methodology applied on different galaxy catalogues \citep{McConnachie+09,DiazGimenez+12}.  

However, the availability of large redshift surveys allows  the CG finder algorithm to be simplified and
improved.  Given that the CG criteria involve the counting of galaxies within a projected area on the sky, not taking into account the position of galaxies along the line of sight 
leads to losing groups by including interlopers, either as galaxy members or as breaches of the isolation.

In our \texttt{modified} algorithm (hereafter \texttt{HICKSON MODIFIED (HM)}), we  therefore consider the redshifts of galaxies from the very beginning of the implementation. Galaxies considered to be neighbours (for membership and isolation) are taken from a cylinder in redshift space around the point where the criteria are being evaluated. Therefore, the velocity filtering is applied at the same time as the other constraints.  
In the implementation of the new algorithm we  also introduce another change to  automatically obtain the largest compact groups in one single run;  in the classic implementation it was run a posteriori. Therefore, we start looking for the largest CGs, and then we select the smaller CG not contained in any previously found larger CG. This is achieved by identifying CGs with the largest numbers of members at first ($N=10$), and then looking for groups with gradually fewer members.  In Fig.~\ref{flux_diagram}, the flow chart that schematises the algorithm is shown on the right.  

As we mentioned above, when the algorithm is applied to a mock lightcone, the blending of galaxies has to be introduced after identifying the CGs to discard those groups whose membership falls below the lower limit when galaxies are blended.

The sample identified in the mock lightcone using this \texttt{modified} algorithm comprises $1287$ \texttt{mHMCG}s (Table~\ref{tab:num}). 

\begin{figure*}
\centering
\includegraphics[width=\textwidth]{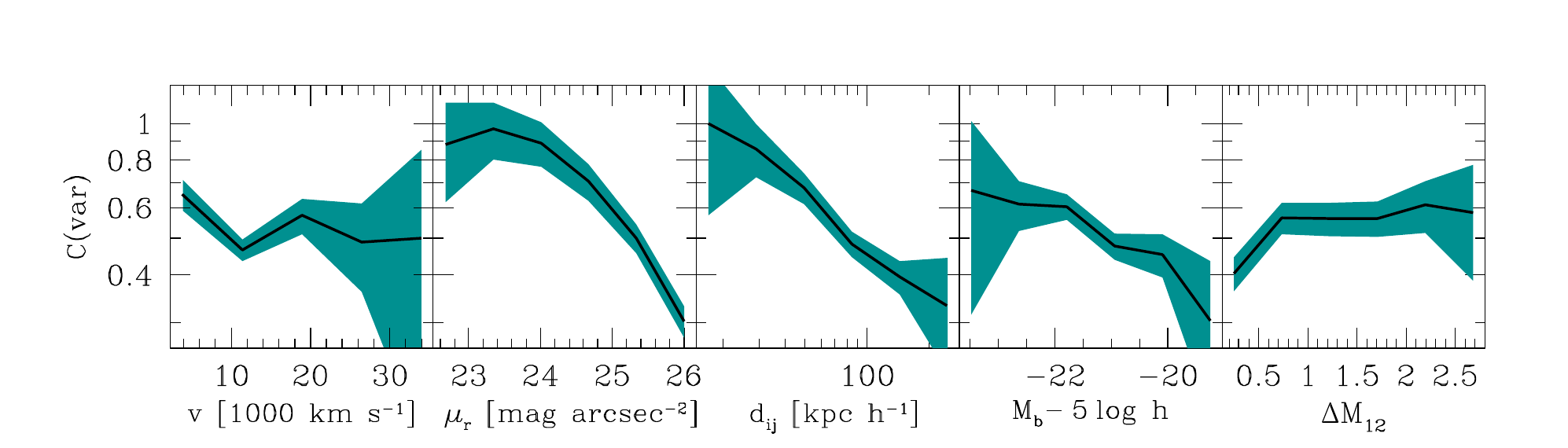}
\caption{Ratio of the classic to the modified distributions for a given property. Shaded areas correspond to the errors computed using the error propagation formula for the ratio (each number in the ratio has a Poisson error).\label{fig:ratio}}
\end{figure*}

\subsection{Comparison of the classic and modified compact group samples}

In this section we compare the two samples of CGs obtained using different algorithms.  
To start with, we note that the \texttt{modified} algorithm produces a sample that is $89\%$ larger than the classic sample. We cross-correlated the two samples and considered that a group from one sample has been recovered in the other sample if the angular distance between their centres are less than the sum of the angular radii of the groups. 
By cross-correlating the samples, we found that $95\%$ of the classic sample (647 \texttt{mHCCG}s) is included in the modified sample.   This means that the new algorithm is able to recover most of the classic sample, but it also identifies almost twice the number of groups, improving the statistics significantly. We also analysed the reasons why the \texttt{classic} algorithm misses almost half of the  \texttt{mHMCG}s. We found that $90 \%$ of the missing groups are discarded by the \texttt{classic} algorithm because there are discordant-velocity galaxies contaminating the isolation annulus in projection. In the remaining $10\%$, the membership was compromised: some of the true group galaxy members were identified in projection, but others were missing, and    new discordant redshift galaxies were  included as group galaxy members. 
These groups did not pass the posterior velocity filtering.

As suggested by \cite{Mamon86} for the HCG sample, 
a non-negligible fraction of mock CGs  are chance projections along the  line of sight \citep{McConnachie+08,DiazGimenez&Mamon10}. It is interesting to examine whether the different algorithms might lead to different percentages of contamination. Following the classification in 3D real space in the simulations performed by \cite{DiazGimenez&Mamon10}, we split the sample into 
real CGs\footnote{Real CGs: 3D comoving maximum interparticle separation between the four closest galaxies less than $100 \, h^{-1} \, \rm kpc$, or less than $200 \, h^{-1} \, \rm  kpc $ while the ratio of  line of sight to transverse size is not higher than $2$.} 
and chance alignments. We found that $(38\pm3)\%$  of the \texttt{mHCCG}s ($256$) are real, while $(35\pm2)\%$  of the \texttt{mHMCG}s ($452$) can be classified as real (Table~\ref{tab:num})\footnote{The percentage of real CGs in this work is significantly smaller than that quoted in previous works. This is being investigated in a different work where different cosmologies and semi-analytic models are considered.}. Therefore, the percentage of physically dense groups is unaffected by the algorithm. 

In Fig.~\ref{fig:boxplot}, we show the distribution of properties of CGs as boxplot diagrams. The first two  boxes in each panel correspond to the mock samples examined in this section. 
The properties shown in these plots are as follows: 
\begin{itemize}
\item ${\rm Redshift}$: group redshift computed as the median of the redshifts of the galaxy members;
\item $\Theta_{\rm G}$: angular diameter of the smallest circumscribed circle; 
\item $\langle d_{ij} \rangle$: median of the inter-galaxy projected separations at the distance of the group centre; 
\item $\mu_r$: $r$-band group surface brightness; 
\item $M_{\rm b}$: $r$-band rest-frame absolute magnitude of the brightest galaxy member;
\item $\Delta M_{12}$: difference in $r$-band rest-frame absolute magnitude between the two brightest galaxies in the group;
\item $\sigma_v$: group velocity dispersion in the line of sight computed using the gapper estimator \citep{Beers+90};
\item $H_0 \, t_{\rm cr}$: dimensionless crossing time, computed as 
$100 \, {h} \pi\langle d_{ij} \rangle /(2 \sqrt{3}\sigma_v);$
\item ${\cal M}_{\rm VT}/L_r$: mass-to-light ratio, where 
$${\cal M}_{\rm VT}=\frac{\pi}{2G}  
R_{\rm VT}^p 
\, 3 \, {\sigma_v}^2,$$ 
$G$ is the gravitational constant, and 
$$R_{\rm VT}^p=
2  \, \langle d_{ij} ^{-1} \rangle^{-1} =2 N (N-1) \left( \displaystyle\sum_j^N\displaystyle\sum_{i \neq j}^N \frac{1}{d_{ij}}\right)^{-1}$$
is twice the harmonic mean projected separation. 
\end{itemize}

In this figure, it can be seen that notches (waists,   related to the confidence intervals) do not overlap for $d_{ij}$, $H_0 \, t_{\rm cr}$, $\Theta_{\rm G}$, $M_{\rm bri}$, and $\mu_r$, which indicates that the medians of these distributions differ significantly \citep{boxplot78,boxplot14}.  The modified sample includes more CGs with larger inter-galaxy projected separations, crossing times, and angular sizes, and fainter first-ranked galaxies and surface brightness. 

To deepen the analysis of whether there is a bias in the classic sample, we measured the completeness of the classic sample as a function of several variables: group radial velocity, group surface brightness, median inter-galaxy projected separations, absolute magnitude of the brightest galaxy, and absolute magnitude gap between the two brightest galaxies. 
We define completeness as the ratio of the number of \texttt{mHCCG}s to \texttt{mHMCG} per bin of each variable ($var$): $C(var)= N_{\texttt{mHCCG}}(var) / N_{\texttt{m HMCG}}(var)$.  The results are shown in Fig.~\ref{fig:ratio}. There is little dependence of the completeness of the classic sample on the radial velocity of the groups: the \texttt{classic} algorithm fails in finding CGs at all distances equally. 
The \texttt{classic} algorithm properly finds the CGs with the brightest surface brightness (lower values of $\mu_r$), but it misses more and more CGs  towards the limit of the compactness criterion. The same happens when analysing the median of the inter-galaxy projected separations: CGs with the largest separations are not detected by the \texttt{classic} finder.  In addition, the \texttt{classic} algorithm misses more CGs with fainter first-ranked galaxies, although it is not even complete at the bright end. Finally, the \texttt{classic} algorithm misses groups in the whole range of magnitude gaps, and the incompleteness is more pronounced for compact groups with two similar first-ranked galaxies.  

\section{Application of the \texttt{modified} compact group finder to the SDSS}
\label{sec:sdss}

In this section we show the implementation of our \texttt{modified} algorithm to the galaxies in the SDSS to build a new catalogue of CGs that strictly meet all Hickson's criteria.

\begin{figure*}
\centering
\includegraphics[width=0.3\textwidth]{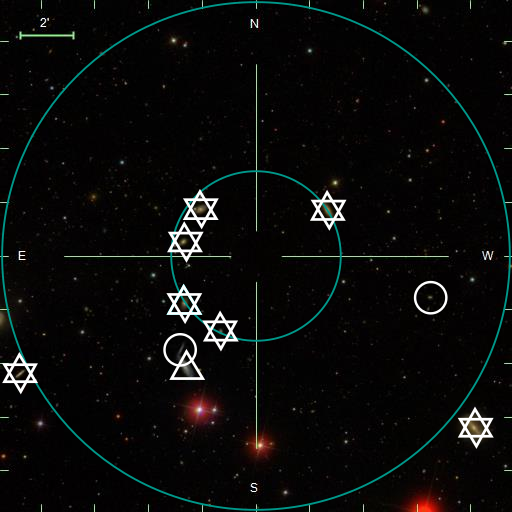}
\includegraphics[width=0.3\textwidth]{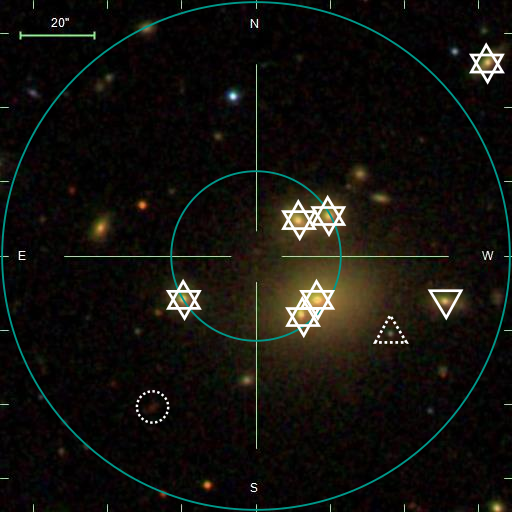}
\includegraphics[width=0.3\textwidth]{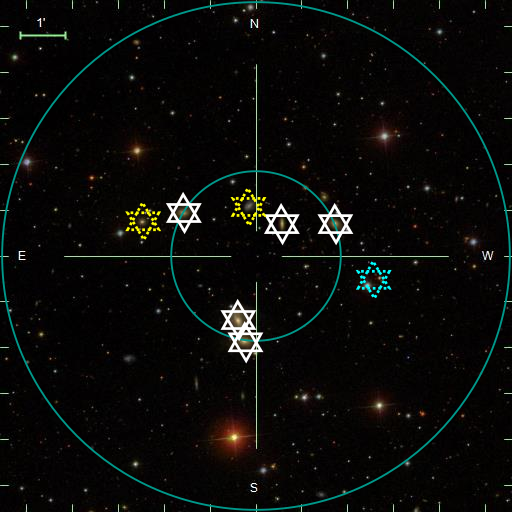}
\caption{Images of three \texttt{HMCG}s in SDSS. 
The inner circles show the minimum circle that encloses all the group members ($\Theta_{\rm G}$), while the outer circles represent the radius of isolation ($3\Theta_{\rm G}$).
Galaxies with spectroscopic information and apparent magnitude $r\le17.77$ are shown  as solid lines, while photometric galaxies with $r\le17.77$ are shown as dashed lines.
Upward-pointing triangles are galaxies in the same magnitude range as the CG members; 
downward-pointing triangles are galaxies in the same redshift range as the CG; 
circles are galaxies outside the magnitude range of the CG. 
Light blue symbols are photometric galaxies considered as non-contaminating (see text), 
while the bright yellow symbols are potentially contaminating objects.
Objects without symbols are either galaxies fainter than $r_{\rm lim}=17.77$ or stars.
Left: Example of an \texttt{HMCG} without any potential sources of contamination (Group ID= 55 in Tab.~\ref{tab:groups}).  Centre: CG with a non-contaminating galaxy within the disc of isolation  whose photometric redshift is clearly outside the redshift range of the CG (Group ID= 210). Right: CG with two potential sources of contamination   $-$ one within the group radius and one in the isolation annulus $-$ that we were not able to discard, while another photometric galaxy in the isolation annulus has been considered as non-contaminating  (Group ID= 460). 
\label{fig:disturbing}
}
\end{figure*}

\subsection{Parent observational galaxy catalogue}
We use the catalogue of galaxies compiled by \cite{tempel17}\footnote{\url{http://cosmodb.to.ee}} from  SDSS Data Release 12 \citep{DR12a,DR12b}. This compilation was constructed using the galaxy spectroscopic information from the main contiguous area of the survey (the Legacy Survey) and it was complemented with redshifts from the Two-degree Field Galaxy Redshift Survey \citep{2df1,2df2}, the Two Micron All Sky Survey Redshift Survey \citep{2mass1,2mass2,2mass3}, and the Third Reference Catalogue of Bright Galaxies \citep{RC3a,RC3b}. Their extended galaxy sample comprises $584 \, 449$ galaxies with observer-frame Petrosian magnitudes $r \leq 17.77$ and redshifts corrected to the CMB rest frame $z_{\rm CMB} \leq 0.2$ within a solid angle of $6828$ square degrees.  In this work, we adopted the model magnitudes as the main apparent magnitudes for the survey. We thus selected galaxies with observer-frame model magnitudes $r$ less than $17.77$ and observer-frame colour $g - r \leq 3$ to avoid stars.  Our final catalogue is formed of  $557 \,
517$ galaxies.   To compute rest-frame absolute magnitudes when needed, k-corrections 
for the $r$ and $g$ magnitudes were computed using the prescriptions of \cite{Chilingarian+12}. 
 The cosmological parameters used here are those obtained by the \cite{Planck+14}.   

\subsection{Compact group extraction}
\label{sec:HMCG}
We applied the \texttt{modified} algorithm (described in Sect.~\ref{modified}) to the galaxies in the extended SDSS catalogue and found a sample of $476$  CGs. 

We visually inspected all the CGs using the SDSS DR14 Image List Tool\footnote{\url{http://skyserver.sdss.org/dr14/en/tools/chart/listinfo.aspx}}. This inspection revealed that $21$ CG members were misclassified as galaxies in the extended sample of \citeauthor{tempel17} when they were actually {Part of a Galaxy} (PofG). The SDSS identification numbers (ObjIDs) of these objects are quoted in Table~\ref{tab:pofg}.  

In addition, it is known that the SDSS spectroscopic survey suffers from incompleteness related to saturated bright galaxies and/or fibre collision for very close projected galaxy pairs. These issues could have an impact on the identification of CGs, mainly in the application of the population and isolation criteria.  Therefore, we used the photometric information of galaxies extracted from the SDSS DR14 \citep{DR14} to search for potential galaxies that are in the surroundings of each identified CG and were neither detected in the spectroscopic survey nor incorporated by \cite{tempel17}. In Appendix~\ref{tablas}, we describe the query used to retrieve galaxies without spectroscopy from SDSS DR14 Casjobs\footnote{\url{http://skyserver.sdss.org/casjobs/}}. 

From the sample of galaxies in SDSS without redshifts, we selected those that lie on the plane of sky within $3\, \Theta_{\rm G}$ of the centre of each identified CG, and whose 
$r$ model magnitudes are within a three-magnitude range from the brightest CG member galaxy ($r \leq r_{\rm b}+3$). This exercise provided us with a list of $303$ objects without redshift information that might contaminate our sample of CGs. 
For these objects, we searched for alternative spectroscopic determinations in the literature using the NASA/IPAC Extragalactic Database (NED)\footnote{\url{https://ned.ipac.caltech.edu/}}.
We found that $63$ of these galaxies had their redshifts already determined from different sources (see Table~\ref{tab:newz}). Objects misclassified as galaxies, as well as missing galaxies, might change the identification of CGs. Therefore, we corrected the parent catalogue of galaxies by discarding objects classified as PofGs and adding galaxies with available redshifts. The \texttt{modified} algorithm was run again on this corrected sample of $557 \, 559$ spectroscopic galaxies.  We obtained a new sample of $462$ \texttt{HMCG} in SDSS which constitutes our final catalogue (see Table~\ref{tab:num}).   

We then again selected $290$ photometric galaxies without redshift measurements that lie on the plane of the sky around the \texttt{HMCG}s in the same way as  described above (angular proximity and magnitude range). 
We examined these objects to classify them
either as potential sources of contamination or `non-contaminating' galaxies. This classification will be useful to determine whether our CGs fulfil the population and isolation criteria. 

Among the $290$ galaxies without SDSS spectroscopic redshifts located in or close to \texttt{HMCG}s, we found that 
\begin{itemize}
\item $52$ galaxies had spectroscopic redshifts as previously identified in \cite{tempel17}; 
\item $53$ other galaxies had photometric redshifts that are clearly discrepant with the \texttt{HMCG} median spectroscopic redshift (following \citealt{Beck+16}, galaxies with $\displaystyle |z_{\rm phot} - z_{\rm cm} |/(1+z_{\rm cm}) > 0.06$ can be safely discarded as outliers);
\item $104$ of the other $185$ galaxies were classified as non-contaminating objects, based on their photometric properties (see Appendix~\ref{nope} for the criterion);
\item $13$ of the remaining $81$ galaxies, after visual inspection via the SDSS DR14 Image List Tool, were determined to have been misclassified as galaxies by the SDSS pipeline;
\item the remaining $68$ objects are  potential sources of contamination.
\end{itemize}

In Fig.~\ref{fig:disturbing}, we show three examples of regions around \texttt{HMCG}s. Pictures were taken from the SDSS DR14 Image Tool. Galaxy members are shown as  upward-pointing triangles combined with downward-pointing triangles (stars) within or on top of the inner circle. The left picture shows an \texttt{HMCG} without any objects with unknown redshift within the group radius or the isolation annulus.
The picture in the centre shows an example where two objects without redshift lie within the isolation radius ($3\, \Theta_{\rm G}$), but 
one of the objects is outside the magnitude range of the group members (open dashed circle), while the other has a photometric redshift that is clearly discordant with the group and therefore  is classified as a non-contaminating galaxy (upward-pointing dashed triangle); the other objects that can be seen in the same field without  symbols are either galaxies fainter than the apparent magnitude limit of the SDSS spectroscopic sample or stars.  Finally, the picture on the right shows a CG that is surrounded by three objects without known redshifts  and in the same magnitude range of the group members,  one galaxy in the isolation annulus that has been classified as non-contaminating based on the method described in  Appendix~\ref{nope} (light blue dashed star), and two others  (one inside the group radius and the other in the isolation annulus) that we were not able to discard as potential contaminants (bright yellow stars).

From the analysis of the potentially contaminating photometric objects from SDSS DR14, 
$18\%$ are spectroscopically confirmed with external spectroscopy and  were already included in our parent catalogue, $59\%$ are clear outliers or misclassified objects, while roughly 23\%\ are uncertain.
Therefore, from our sample of $462$ \texttt{HMCG}s, 406 of them can be considered clean (i.e. without sources of contamination), while  56 \texttt{HMCG}s need further information of galaxies in their surroundings in order to discard potential contamination sources.  

In Appendix~\ref{catalogue}, we present the new catalogue of \texttt{HMCG}s. Groups with potential contamination have not been discarded from our sample, but they are presented at the end of the tables with a flag that indicates the need for spectroscopic information of the potential sources of contamination (see Table~\ref{tab:intruder}). 

For the sake of comparison with the results found from the mock catalogues, we applied the \texttt{classic} algorithm to the extended and corrected sample of galaxies from SDSS, and we obtained a sample of 218 \texttt{HCCG}s (see Table \ref{tab:num}). Similar to the results found in the mock catalogue, the \texttt{modified} algorithm was able to recover $95\%$ of these groups, while it nearly doubles the number of CGs identified. Also, the \texttt{classic }algorithm misses almost half of the \texttt{HMCGs}  due to discordant-velocity galaxies contaminating the isolation annulus.
In Figure \ref{fig:boxplot}, we show the boxplots obtained for the properties of CGs identified using both algorithms, where the samples are labelled  Hickson classic (\texttt{HC}) and  Hickson modified (\texttt{HM}). 
From this comparison, we observe  trends similar to those in the samples obtained from the mock catalogue: the median values of $d_{ij}$, $H_0 \, t_{\rm cr}$, $\Theta_{\rm G}$, and $\mu_r$ are significantly higher with the \texttt{modified} method than with the \texttt{classic} method. Moreover, the analyses of completeness as a function of different properties produced the same results found for the mock samples. 
The completeness of the sample of CGs has therefore been improved with the new implementation of the algorithm.

\subsection{Comparison with other compact group catalogues from SDSS}
A number of samples of CGs extracted from the SDSS have been presented in the literature. Particularly, \cite{McConnachie+09}  (hereafter \texttt{McC09}) identified CGs in projection in the SDSS Data Release 6 (DR6).  
It is worth mentioning that: 
\begin{enumerate}[i)]
\item The identification in projection done by \texttt{McC09} 
was performed up to magnitudes fainter than the SDSS spectroscopic limit ($r_{\rm lim}=17.77$). One of their samples (catalogue A) is identified down to $r \le 18$, while the other (catalogue B) is for $r\le 21$.
\item \texttt{McC09}
did not take into account the flux limit criterion ($r_b\le r_{lim}-3$). Therefore, the isolation and population criteria were not strictly checked (as mentioned by \texttt{McC09}). Moreover, the lack of flux limit criterion could introduce an artificial correlation between group size and redshift.
\item The CG identification was performed on a subsample of SDSS DR6 with only galaxies fainter than $r_{\rm inf}=14.5$. This restriction led to a sample that misses low redshift CGs. 
\end{enumerate}

Later, \cite{sohn+15} looked for velocity information from different sources for the galaxy members of \texttt{McC09} CGs, and applied a velocity filter afterwards (\texttt{SOHN15} CGs).
In the best-case scenario, the resulting sample would be equivalent to the implementation of our \texttt{classic} algorithm.  

Another CG catalogue was extracted from SDSS DR12 by \cite{sohn+16}. They used a percolation algorithm with two fixed linking lengths, and looked for systems with high overdensity contrast (small intergalaxy separations). They explicitly disregarded the isolation criterion, which could lead to a sample of small substructures within larger systems. This catalogue will be  examined further in the following subsection.

In Table~\ref{tab:cat}, we show the main characteristics of these CG samples extracted from different releases of SDSS\footnote{The CG samples used in this work were downloaded from \url{http://vizier.u-strasbg.fr/viz-bin/VizieR}}. The first six rows refer to the selection criteria of each sample. The middle rows have the number of CGs in each sample.  Each row in this part of the table is obtained applying the restriction indicated in the first column and including the restrictions of the previous rows. Finally, the last four rows show the median of some properties of the samples of CGs resulting after applying all the restrictions to make the samples comparable. Given the small number of CGs in the restricted \texttt{McC09} CG samples, these values cannot be included.
\begin{table}
\caption{Catalogues of compact groups from SDSS. \label{tab:cat}}
\begin{center}
\small
\setlength{\tabcolsep}{2pt}
\begin{tabular}{rrrrrr}
\hline
\hline
 & \multicolumn{2}{c}{\texttt{McC09}} & \texttt{SOHN15} & \texttt{SOHNCG} & \texttt{HMCG}\\
\cline{2-3}
& A \ \ & B \ \ \\
\hline
Release & DR6 & DR6 & DR12$^+$ & DR12$^+$ & DR12$^+$ \\
 finder & PRO & PRO & P+VF & FoF & MOD \\ 
$r_{\rm lim}$ & 18 & 21 & 17.77 & 17.77 & 17.77\\
$r_{\rm inf}$ & 14.5 & 14.5 & $--$ & $--$& $--$\\
$\mu_{\rm lim}$ & 26 & 26 & 26 & 26 & 26.33 \\
isolation & YES & YES & YES & NO & YES \\
 
\hline
Total & 2297 & 74791 & 332 & 1588 & 462 \\
$N \ge 4$ & 2297 & 74788 & 192 & 312 & 462 \\
+ $N_z=N$ & 153 & 55 & 140 & 312 & 462 \\ 
+  $\Delta {\rm v}_{i,{\rm cm}}\le 1000 $& 44 & 13 & 140 &312 & 462\\
+ $r_{\rm b} \le 14.77$ & 2 & 2 & 11 & 142 & 462\\
+ $\mu_r \le 26$ & 2 & 2 & 11 & 142 & 383 \\
\hline
$v \, [10^3 \, \rm km \, s^{-1}]$ & $--$&$--$ & $19.6(2.7)$& $10.6(0.7)$& $12.2(0.6)$\\
$r_p \, [h^{-1} \rm kpc]$& $--$ &$--$ & $40(8)$ & $ 38(2)$ & $73(3)$ \\
$\sigma_v \, [\rm km \, s^{-1}]$ & $--$&$--$ & $215(71)$ & $323(37)$ & $236(16) $\\
$\Delta \rm M_{12}$ &$--$ &$--$ & $1.4(0.4)$ & $0.8(0.1)$ & $ 1.0(0.1)$ \\
\hline
\hline

\end{tabular}
\parbox{0.45\textwidth}{Notes. Compact group catalogues: \texttt{McC09}=\cite{McConnachie+09}, \texttt{SOHN15}=\cite{sohn+15}, \texttt{SOHNCG}=\cite{sohn+16}.
Releases: the $+$ upper sign means that the data were complemented with redshift information from different sources. 
Finders: PRO = in projection, P+VF= in projection plus posterior velocity filtering, FoF = friends-of-friends algorithm, MOD = \texttt{modified} algorithm (this work).
In the middle section of the table, $N_z$ is the number of members with spectroscopic information, while the $+$ sign at the beginning of each row indicates that it includes the restrictions of the previous ones.  
In the four bottom rows, the format xx (ss) contains the median of the properties (xx) and the shift (ss) to construct an approximated 95\% confidence interval, $CI = xx \pm ss$, where $ss=1.58 \times {\rm IQR} /\sqrt[]{n}$,  $n$ is the number of objects in the sample and IQR is the interquartile range \citep{boxplot14}.
}
\end{center}
\end{table}

\begin{figure*}
\centering
\includegraphics[width=.8\textwidth,trim=0 0 50 0,clip]{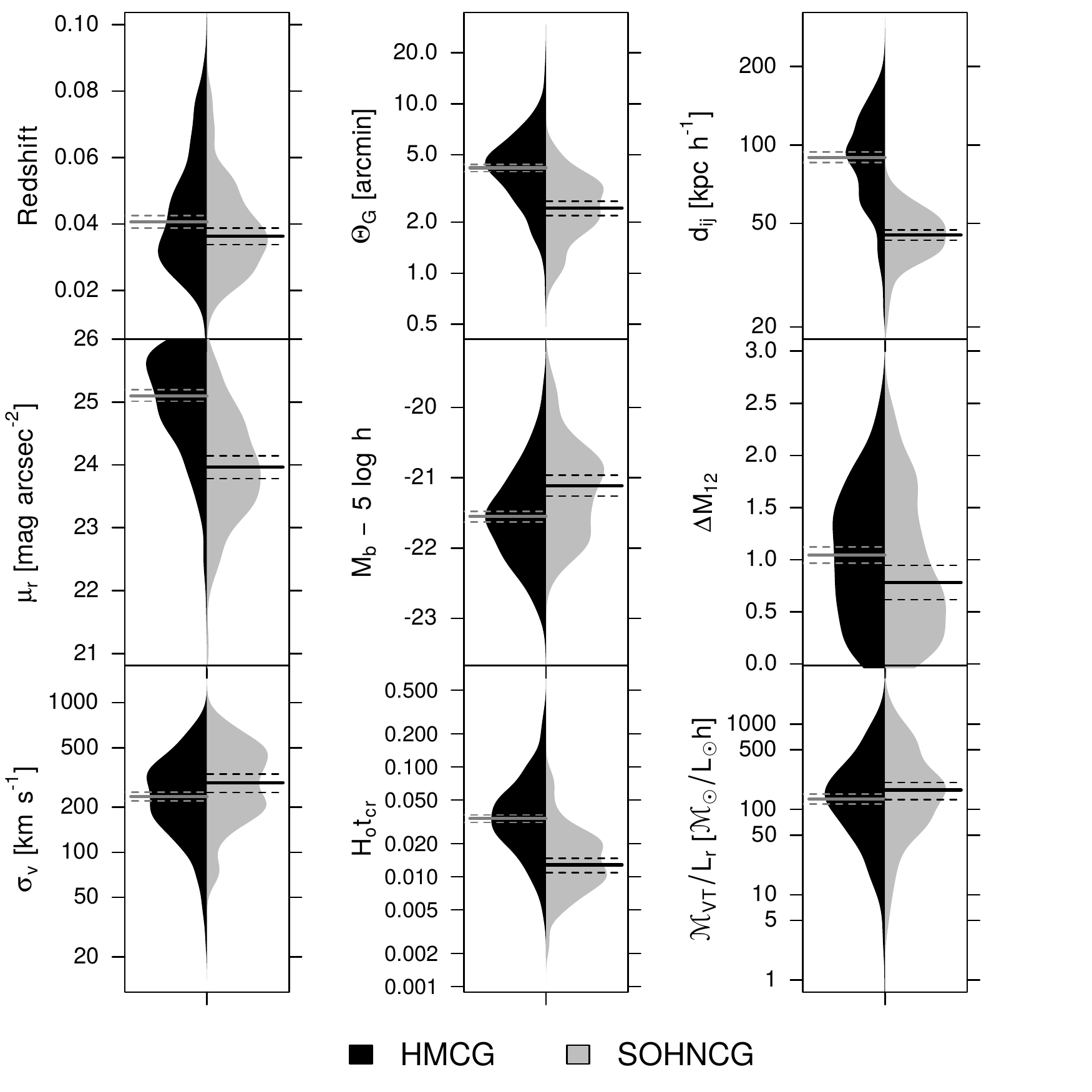}
\caption{Comparison of CG properties using asymmetric beanplots \citep{beanplot+08}. The black beans represent the density distributions for CGs identified with the \texttt{modified} algorithm (\texttt{HMCGs}), while grey distributions corresponds to the CGs that belong to the \texttt{SOHNCGs} sample. Both samples were restricted to perform a fair comparison (see text for details). Horizontal solid lines show the median values for each property distribution, while the horizontal dashed lines show the upper and lower limits of their corresponding $95\%$ confidence intervals \citep{boxplot14}. This figure was made with R software \citep{plotR}.\label{fig:beanplot}}
\end{figure*}

\subsubsection{ \texttt{HMCG}s vs \texttt{SOHNCG}s}
As mentioned before, it is always difficult to compare samples of CGs identified by different authors given the inherent differences between the parent catalogues, the photometric bands, the algorithms,  the definitions of what a compact group is, etc. However, we embraced this task and performed a detailed comparison between the new observational \texttt{HMCG} sample described in Sect.~\ref{sec:HMCG}  and the CG sample identified by \cite{sohn+16}. 

Similar to our work, \citeauthor{sohn+16} have also used the catalogue of spectroscopic galaxies from the SDSS DR12 \citep{DR12b}, but they complemented the survey themselves adding spectroscopic information extracted from other surveys such as \cite{hwang+10} and NED, especially for bright galaxies ($r<14.5$). Their final flux-limited galaxy sample includes $654 \, 066$ galaxies with magnitudes $r<17.77$ in the redshift range $0.01<z<0.20$.

Unlike us, the identification of CGs made by \cite{sohn+16} was performed using a FoF  algorithm similar to that used by \cite{Barton96}. Their FoF algorithm is based on two fixed linking lengths that restrict the projected physical spatial distance between galaxies ($\Delta D \leq 50$ ${\rm kpc \ h^{-1}}$) and the rest-frame line-of-sight velocity separation ($\Delta V\leq$ 1000 ${\rm km/s}$) of every pair of neighbours. Moreover, in the spirit of the Hickson compact groups, they included the population and compactness criteria ($N(\Delta r \leq 3) \geq 3$ and $\mu_r \leq 26 \ {\rm mag \ arcsec^{-2}}$). However, they intentionally avoided  applying any isolation criteria to their CGs.
They presented a sample of 1588 CGs\footnote{Publicly available as table \url{http://vizier.u-strasbg.fr/viz-bin/VizieR-3?-source=J/ApJS/225/23} in VizieR}. 
To perform a fair comparison with our sample, we needed to include several restrictions. 
To start with, we selected from their sample only those CGs with four or more galaxy members. The restricted sample comprises 312 CGs. Furthermore, we selected those groups that meet the flux limit criterion ($r_{\rm b}\leq r_{\rm lim}-3$) for which the population criterion has fully been checked, leading us to a smaller sample of 142 CGs. 
Finally, we applied the sky angular mask of the \cite{tempel17} catalogue to have the same angular coverage. All in all, the final sample comprises $109$ groups  (hereafter \texttt{SOHNCG}).  By angularly cross-correlating this sample and our \texttt{HMCG}s, we found that $ 44 \%$ of the $109$ \texttt{SOHNCG}s are included in our sample. The reasons why we were not able to recover the remaining $56 \%$ are discussed below.

In addition, to compare both samples we  also restricted our own sample to mimic the surface brightness limit imposed by \citeauthor{sohn+16} The two limits are different since they have kept the original limit from Hickson's criterion in the $R$-band ($\mu_R < 26$), while we have converted the limit to the $r$-band ($\mu_r < 26.33$). With $\mu_r < 26.0$, our sample decreases from $462$ to $383$ \texttt{MHCG}s.  After making these samples comparable,  we note that our \texttt{HMCG} sample is a factor of 3.5 larger than \texttt{SOHNCG}. 

\begin{figure*}
\centering
\includegraphics[width=0.45\textwidth]{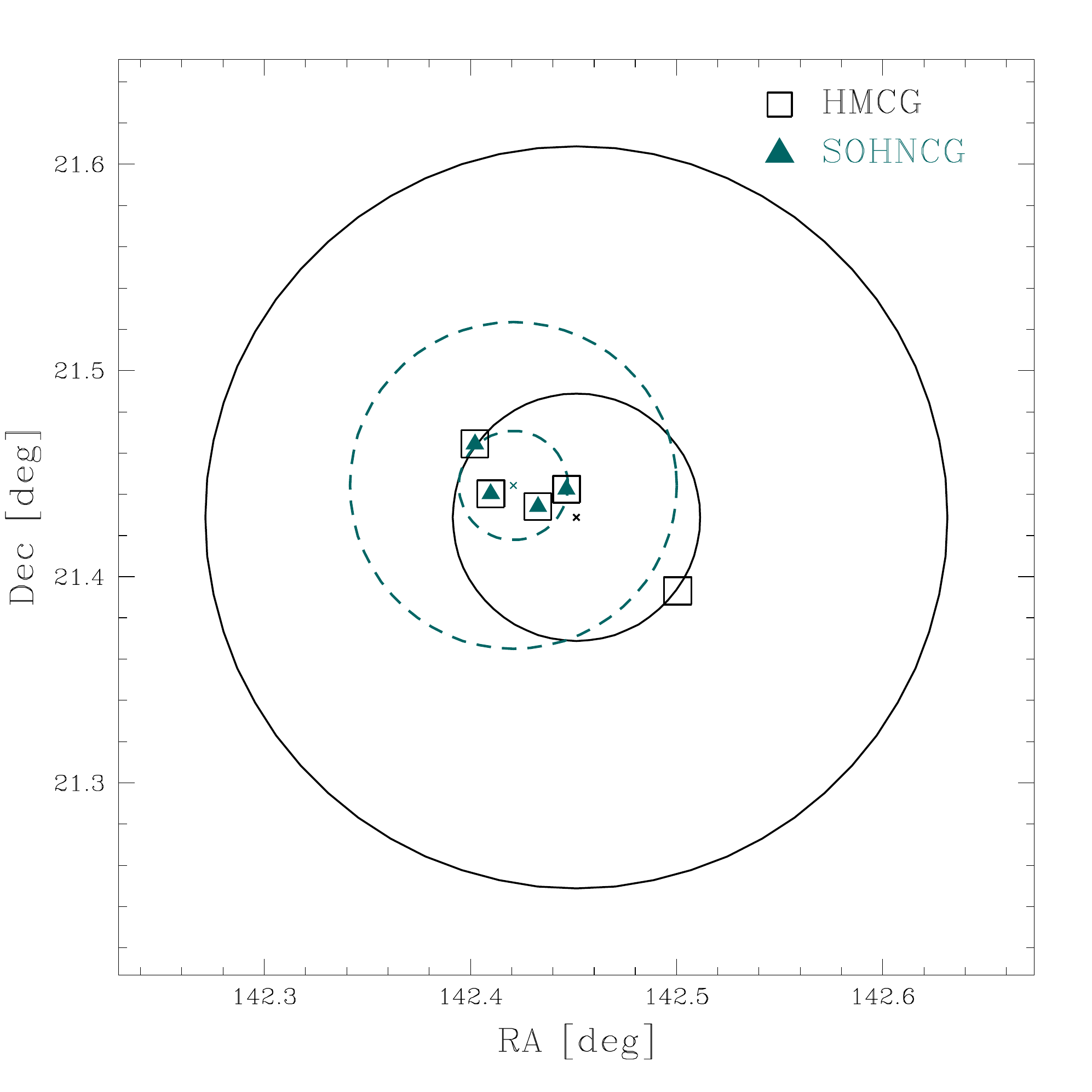}
\includegraphics[width=0.45\textwidth]{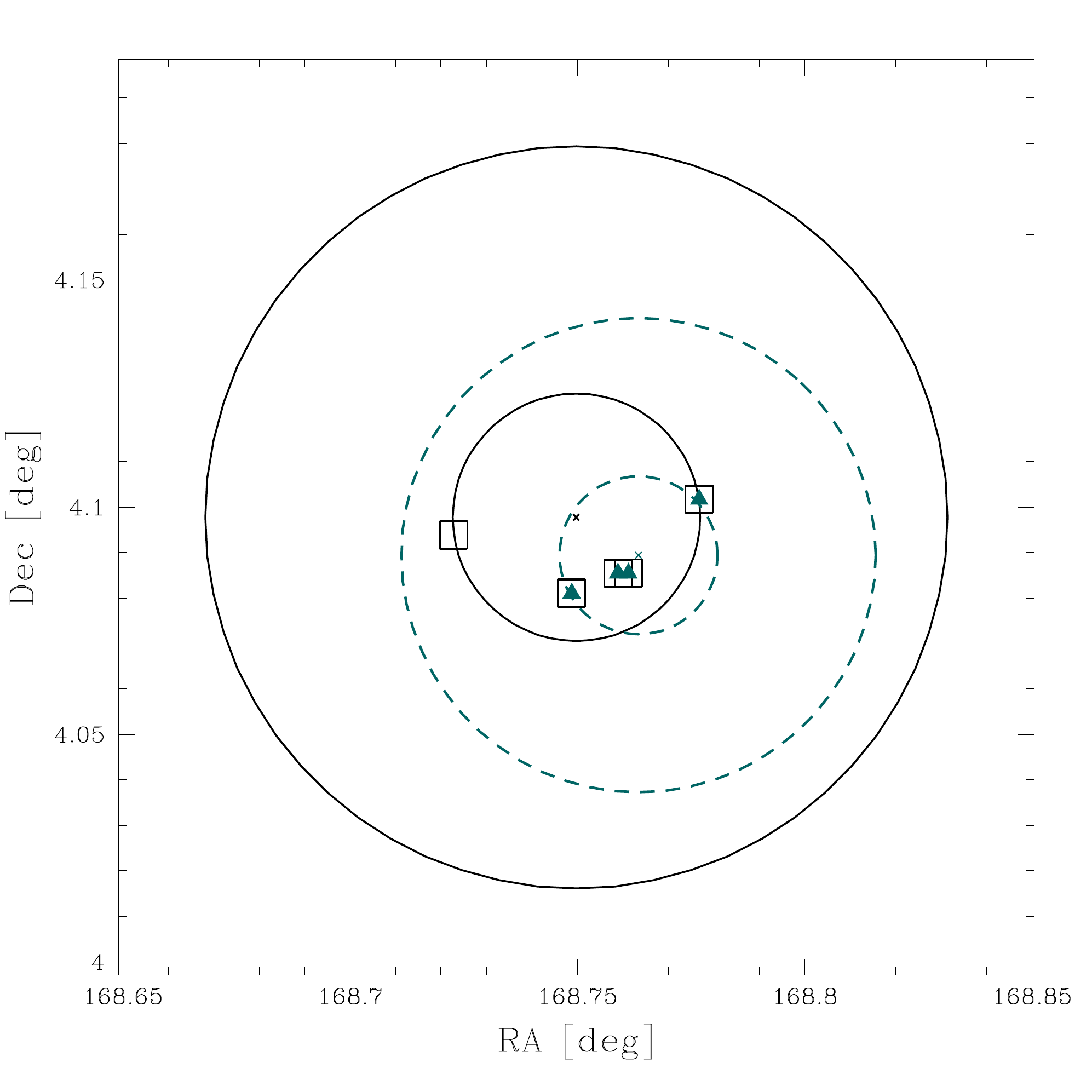}
\caption{Compact groups  in common in the \texttt{HMCG}s and \texttt{SOHNCG}s samples. The \texttt{HMCG} galaxy members are represented as black squares, while the \texttt{SOHNCG} members are cyan triangles. The inner circle shows the minimum circle that encloses all the group members ($\Theta_{\rm G}$), while the outer circle represents the disc of isolation ($3\Theta_{\rm G}$). Each group centre is represented by a dot. 
Left panel: Example of an embedded CG. The \texttt{SOHNCG} (Group ID= 600) lies inside the \texttt{HMCG} (Group ID= 53), while there are no galaxies within the disc of isolation of the \texttt{SOHNCG}. Right panel: Example of a non-isolated subgroup (\texttt{SOHNCG} ID= 1115) that lies inside an \texttt{HMCG} (Group ID = 137).  
\label{fig:recuperados}
}
\end{figure*}

In Figure~\ref{fig:beanplot}, we compare the properties of these two samples of SDSS CGs. In this figure, we show asymmetric beanplots, i.e. the averaged density shape with a proper normalisation (see \citealt{beanplot+08}) for a given property, allowing us to make a direct comparison among the samples. Horizontal solid lines represent the median of the given property, while dashed lines represent their confidence intervals.  
From this comparison, we observe that the main difference is that \texttt{SOHNCG}s are  smaller, producing $\Theta_{\rm G}$, $d_{ij}$, $\mu_r$, and $H_0 \, t_{\rm cr}$ distributions that are shifted towards lower values compared with the values obtained for our \texttt{HMCG} sample. These results could be related to the identification procedure: the small projected linking length used in the percolation algorithm by \cite{sohn+16} leads to groups that are smaller in projection than those defined via Hickson-like finders. 
In addition, we observe that \texttt{HMCG}s show statistically brighter first-ranked galaxies 
and larger magnitude gaps between the two brightest galaxies  ($\Delta M_{12}$) than their counterparts in \texttt{SOHNCG}s. Therefore, our sample is more dominated by its brightest galaxies. 

Moreover, even though the median redshift for our \texttt{HMCG} sample is slightly shifted towards higher values (a difference of $\sim 0.005$ in redshift or $\sim 1500 \ {\rm km \ s^{-1}}$) compared with the corresponding  \texttt{SOHNCG}s, this difference is not statistically significant (both lie within the 95\% confidence interval of the other median) and the two density distributions are very similar. 
In comparison, \cite{sohn+16} found that the redshift distribution of their CG sample was roughly $9000 \ {\rm km \ s^{-1}}$ lower (closer) than that of the sample that \cite{McConnachie+09} identified in the SDSS DR6 (with velocity filtering subsequently performed by \citealt{sohn+15}). \cite{sohn+16} attributed this difference to the isolation criterion incorporated by \cite{McConnachie+09}. However, our \texttt{classic} and \texttt{modified} CGs have a median redshift similar to the \cite{sohn+16} values.
It should be noted that \cite{sohn+16} performed a comparison between samples restricted to groups whose brightest galaxy was fainter than $14.5$ because \cite{McConnachie+09} identified their CGs with this restriction. 
This omission of very bright galaxies might be a better explanation of the different redshift ranges of the \cite{McConnachie+09} and \cite{sohn+16} CG samples  than the inclusion or not of the isolation criterion.

Finally, to deepen our understanding of the differences found between our \texttt{HMCG}s and \texttt{SOHNCG}s, we first restricted the analysis to the $49$ groups that are in common in both samples. Given that the matching has not been done on a one-by-one member basis, groups that are in common might still differ in their properties. In fact, we found that the differences mentioned above hold even for the samples that contain only the groups present in both samples. The \texttt{HMCG}s could be expected to be larger in projection because we intentionally selected them in this way: in the cases where there are galaxy associations that meet  all the CG criteria inside larger associations that also fulfil all the criteria (embedded CGs), we preferred to keep the largest groups. To disentangle this issue, we then performed a  member-by-member comparison and discovered that only seven cases (see left panel in Fig.~\ref{fig:recuperados}) are embedded CGs in our sample. Also, we observe that there are 12 \texttt{SOHNCG}s that are subgroups inside our \texttt{HMCG}s that are not picked as CGs in our algorithm because the isolation is broken if the other galaxies are not included as part of the groups (see right panel of Fig.~\ref{fig:recuperados}). Finally, there are nine groups where  half of the galaxy members are not shared, even when the centres are very close, because the brightest galaxy is different and therefore the magnitude range in which the neighbours are considered differs. The remaining 21 groups are exactly the same. In the first two cases (which constitute $\sim 40\%$ of the sample) the \texttt{SOHNCG}s are basically smaller substructures inside our groups which lead to the differences in sizes that we reported before. So, we conclude that the smaller sizes of \texttt{SOHNCG}s could be a consequence of the different definition of what a compact group is for each author more than a result of  the algorithms themselves. 
On the other hand, analysing the $60$ \texttt{SOHNCG}s that we were not able to recover, we found that we missed them because they failed our isolation criterion, while \cite{sohn+16} did not apply this criterion.  

\section{Summary}
\label{sec:discus}
The original definition of compact groups states that they are small, high density associations of bright galaxies that are relatively isolated in space \citep{Hickson82}. To identify such systems, groups must meet several criteria:
limited population within a magnitude range, compactness, spatial isolation within a magnitude range, and velocity concordance of all of their galaxy members. While in principle the limiting values that can be adopted for each of the criteria are arbitrary, it is customary to adopt the definitions introduced by \cite{Hickson82} and \cite{Hickson92}. In this work, we followed these original ideas, and adopted the commonly used limits to present a new algorithm to identify CGs in redshift surveys. 

The algorithm has been tested using a mock lightcone built from galaxies extracted from a recent semi-analytical model of galaxy formation \citep{Henriques+15} run on top of the Millennium Simulation I rescaled to represent the cosmological model determined by the Plank cosmology. Given the methodology used to test the algorithm, the choice of a different semi-analytical model will not affect the results presented in this work. 

To test the new algorithm we compared the resulting sample with that obtained using a previous version of the algorithm that had been used in the past to create and analyse samples of CGs \citep{DiazGimenez&Mamon10,DiazGimenez+12,DiazGimenez+15,Taverna+16}. 
We called the previous application of the algorithm  \texttt{classic}, since it basically reproduces the  steps followed by \citeauthor{Hickson82}: it starts looking for CG in projection in the plane of the sky, and only when the projected sample is available does it check whether all the galaxy members have concordant radial velocities. Other authors have also followed these steps \citep{MendesdeOliveira&Hickson91,McConnachie+09,sohn+15}. The main disadvantage of the \texttt{classic} path is that when the algorithms involve the counting of galaxies in a magnitude range within a region of the projected sky, it may include many interlopers that lie in the line of sight in the background or foreground of the group itself. These interlopers will affect the population, as well as the isolation of the potential groups, and eventually some of the groups may be excluded from the projected sample. This could thus compromise the completeness of the resulting sample. 

With the issue of the completeness in mind, we introduced a \texttt{modified} version of the algorithm. Instead of applying the velocity filter on a pre-selected projected sample of CGs, the velocity concordance is a requirement imposed as the galaxies are being included as galaxy members or neighbours. 

As a result, we obtained modified samples that are roughly twice the size of the classic samples. We have shown that, compared to the \texttt{classic} version, the \texttt{modified} searching of CGs includes more  groups towards the limit of the compactness criterion, with larger projected sizes, with fainter first-ranked galaxies, and with more similar two first-ranked galaxies. The incompleteness of the classic sample shows no dependence on distances to the groups. 

Therefore, our \texttt{modified} algorithm helped us to improve the completeness of the samples of CGs. An interesting question now arises about the purity of the samples. Using the information in 3D real space from the simulation, and following \cite{DiazGimenez&Mamon10}, we investigated the occurrence of chance alignments in our classic and modified CG samples. We found that 
the algorithm does not affect the fraction of CGs that are chance alignments of galaxies along the line of sight.
It would have been desirable to diminish the fraction of chance alignment groups, but at least it has not increased even though we  doubled the number of identified CGs. In a future work, we will investigate how to increase the fraction of physically dense compact groups (i.e. diminish the fraction of chance alignment groups) via observational constraints. 

As a corollary of this study, we present a new CG catalogue identified on the observational spectroscopic galaxy sample compiled by \cite{tempel17} from the SDSS DR12. This CG catalogue comprises $462$ systems. After performing visual and automatic inspections of the surroundings of each CG in the photometric SDSS sample, we determined that $406$ CGs ($\sim 88\%$ of the sample) can certainly be considered isolated, while for the remaining $56$ CGs we were not able to classify them as surely isolated, due to possible photometric contamination in their surrounding area caused by incompleteness inherent to the SDSS spectroscopic catalogue (saturation of bright galaxies and fibre collision). Further spectroscopic information is needed. 

In addition, we performed a detailed comparison with the available sample of CGs identified by \cite{sohn+16}. The comparison between samples of CGs identified by different authors, algorithms, and criteria is never straightforward, and one has to be careful when extracting general conclusions about compact groups that could stand for CGs defined in one way but do not hold for a different sample. It is important to have larger samples of CGs identified in a unique homogeneous way, with well known selection criteria to obtain statistically reliable conclusions. With this aim either the sample of \cite{sohn+16} and the sample introduced in this work achieve this goal, and  to shed light on the properties of small peculiar galaxy systems, although the two samples differ in their definitions of what the authors understand by compact groups. Particularly, criteria regarding flux limit, isolation and embedded compact groups are key in determining the properties of the final sample of CGs. 

The sample of 462 Hickson Modified Compact Groups (\texttt{HMCG}s) presented in this work has become the largest  catalogue of CGs that satisfy all of Hickson's original criteria: they are small, compact, and isolated associations of four or more concordant galaxies. This sample is available for the astronomical community as electronic table in this publication (see Appendix~\ref{catalogue} for details).

\section*{Acknowledgements}
{\small  The authors would like to thank the referee, Dr. Gary Mamon, for his  thoughtful analysis and suggestions that helped us to improve the original manuscript.  

The Millennium Simulation databases used in this paper and the web application providing online access to them were constructed as part of the activities of the German Astrophysical Virtual Observatory (GAVO).

The observational data, which is a recompiled version of the publicly available SDSS data, was extracted from \url{http://cosmodb.to.ee/}. 
Funding for SDSS-III has been provided by the Alfred P. Sloan Foundation, the Participating Institutions, the National Science Foundation, and the U.S. Department of Energy Office of Science. The SDSS-III website is \url{http://www.sdss3.org/}. SDSS-III is managed by the Astrophysical Research Consortium for the Participating Institutions of the SDSS-III Collaboration including the University of Arizona, the Brazilian Participation Group, Brookhaven National Laboratory, Carnegie Mellon University, University of Florida, the French Participation Group, the German Participation Group, Harvard University, the Instituto de Astrofisica de Canarias, the Michigan State/Notre Dame/JINA Participation Group, Johns Hopkins University, Lawrence Berkeley National Laboratory, Max Planck Institute for Astrophysics, Max Planck Institute for Extraterrestrial Physics, New Mexico State University, New York University, The Ohio State University, Pennsylvania State University, University of Portsmouth, Princeton University, the Spanish Participation Group, University of Tokyo, University of Utah, Vanderbilt University, University of Virginia, University of Washington, and Yale University. This research has made use of the NASA/IPAC Extragalactic Database (NED) which is operated by the Jet Propulsion Laboratory, California Institute of Technology, under contract with the National Aeronautics and Space Administration. 

This work has been partially supported by Consejo Nacional de Investigaciones Cient\'\i ficas y T\'ecnicas de la Rep\'ublica Argentina (CONICET) and the Secretar\'\i a de Ciencia y Tecnolog\'\i a de la Universidad de C\'ordoba (SeCyT)}

\bibliography{biblio}

\begin{appendix}
\section{K-decorrection}
\label{k-corr}
To obtain the observer-frame magnitudes for galaxies in the lightcones, one has to `k-decorrect' the rest-frame magnitudes provided by the SAMs. For this purpose, we developed an iterative process based on observational data. 

\begin{figure}
\begin{center}
\includegraphics[width=\hsize]{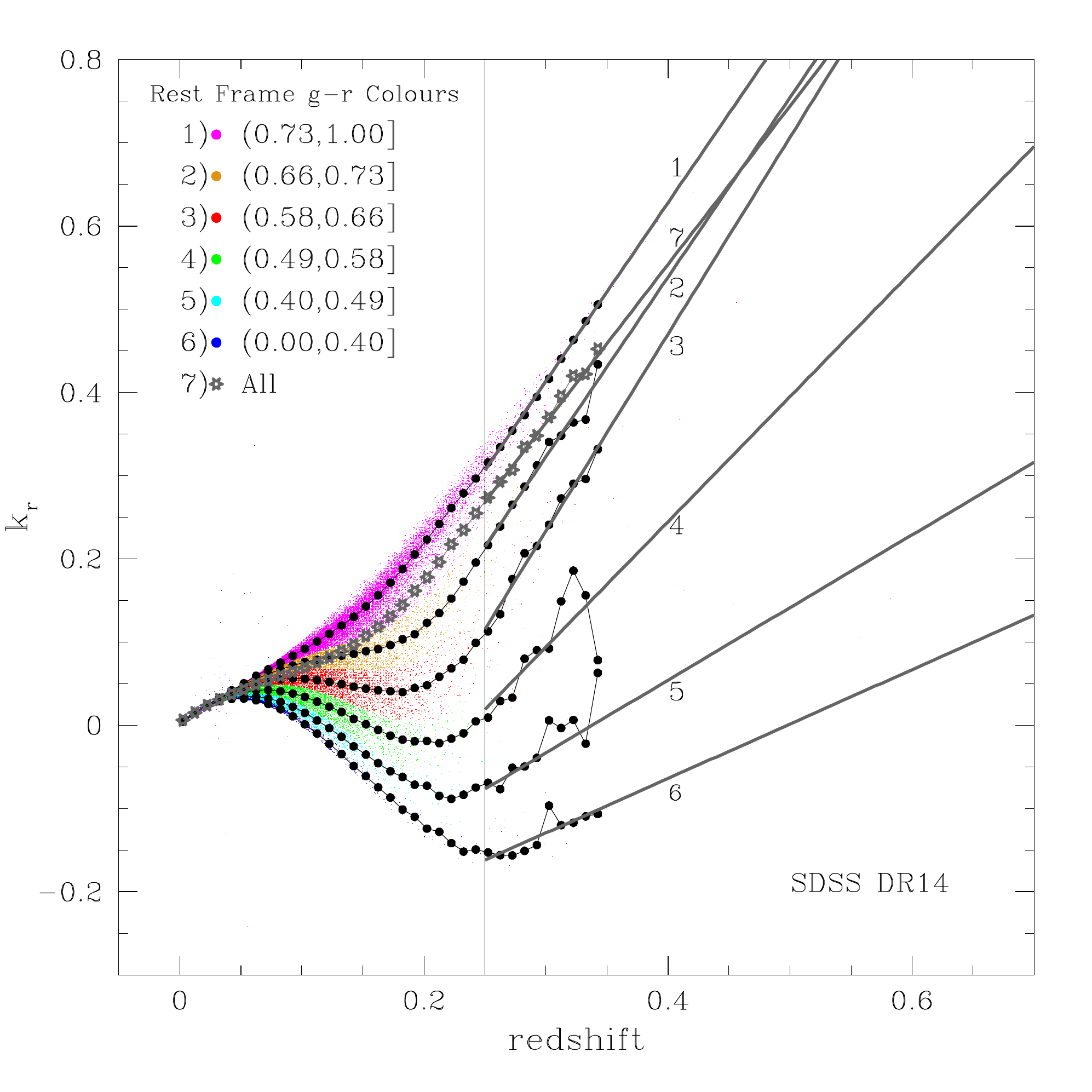}
\caption{\label{f6} k-correction vs redshift for galaxies in the SDSS DR14 split by rest-frame colours (see inset labels). 
k-corrections are computed using \citealt{Chilingarian+12}.
Black dots are the median values of the k-corrections for a given redshift bin for each colour range. Grey stars are the medians for the whole sample of galaxies regardless their colour. Black straight lines are the linear fits to the dots with redshift greater than 0.25 (vertical solid line). 
} 
\end{center}
\end{figure}

First, using the observer-frame magnitudes and colours of galaxies in the SDSS DR14
we computed the k-corrections for those galaxies using the prescriptions of \cite{Chilingarian+12} which, in addition to redshifts, use the observer-frame magnitude and colours as inputs. Figure~\ref{f6} shows the k-corrections in the $r$-band as a function of redshift. Once the k-corrections are obtained, we computed the rest-frame colours of galaxies, and split the sample into several colour ranges. For each rest-frame colour range, we computed the median values of k-corrections per bin of redshift. 
We considered these median values as a good approximation up to a limiting distance ($z=0.25$ in Fig.~\ref{f6}).  
For larger values of redshifts, we computed a linear fit for each rest-frame colour range. For galaxy colours outside the ranges defined in the figure (i.e. $g-r\le 0$ OR $g-r>1$), we adopt the values corresponding to the total sample of galaxies (grey stars in the figure). 

In this way, using the observational fits, we are able to obtain a first approximation to the k-corrections of galaxies when we only have access to their rest-frame colours and redshifts: for galaxies at redshifts lower than $z=0.25$, we use linear interpolations between the median values of k-corrections per bin of redshifts (black dots in Fig.~\ref{f6}). For higher redshifts, we used the linear fits obtained from the data\footnote{For galaxies with g-r colour less than 0 or greater than 1, we adopted the medians and the fit obtained for the whole sample of galaxies (yellow dots in Fig.~\ref{f6})}.  Then, we compute the observer-frame colours and use them as inputs in the original \citeauthor{Chilingarian+12} prescriptions.
With these k-corrections, we then
compute new observer-frame magnitudes and iterate into the \citeauthor{Chilingarian+12} prescriptions one more time to compute final k-corrections to k-decorrect the original rest-frame magnitudes and estimate the final observer-frame magnitudes. 

We tested this iterative procedure using SDSS DR14 galaxies. Starting with the observer-frame magnitudes from the catalogue, we obtained the k-corrections via \citeauthor{Chilingarian+12} and computed their rest-frame magnitudes. These rest-frame magnitudes are used as inputs in our iterative process in which the observer-frame magnitudes are estimated.  Comparing the {\it observed} observer-frame magnitudes with the {\it estimated} observer-frame magnitudes, we obtained that the $99th$ percentile of the difference of $| r_{\rm observed} - r_{\rm computed}| $ is $0.009$ magnitudes, which indicates the success of the method. 

The Fortran subroutine used to k-decorrect magnitudes and obtain observer-frame magnitudes in mock catalogues starting from rest-frame magnitudes is publicly available at the author's institute web page\footnote{\url{https://iate.oac.uncor.edu/index.php/alcance-publico/codigos/}}. The code can also be used for the $K_{2MASS}$-band which has been tested on galaxies from the 2MRS catalogue \citep{2MRS}.

\section{Tables of galaxies}
\label{tablas}
In this appendix we provide further information about galaxies used in this work:
\begin{itemize}
\item In Table~\ref{tab:pofg} we quote the SDSS object identification number (ObjID) of $21$ galaxies in the \cite{tempel17} catalogue that were visually classified as {part of a galaxy} (PofG).

\item We selected galaxies from SDSS DR14 for which the SDSS-III imaging pipeline has declared the photometry clean\footnote{\url{http://skyserver.sdss.org/dr14/en/help/docs/realquery.aspx\#cleanGals}}. The query is quoted below:

\begin{lstlisting}[
           language=SQL,
           breaklines=true,           
           showspaces=false,
           basicstyle=\ttfamily,
           numbers=left,
           numberstyle=\tiny,
           commentstyle=\color{gray}
        ]
SELECT px.objID,px.ra,px.dec,px.modelMag_u,px.modelMag_g,px.modelMag_r,px.modelMag_i,px.modelMag_z,px.modelMagErr_u,px.modelMagErr_g,px.modelMagErr_r,px.modelMagErr_i,px.modelMagErr_z,px.extinction_u,px.extinction_g,px.extinction_r,px.extinction_i,px.extinction_z,pxx.petroRad_r,pxx.petroRadErr_r,pxx.petroR50_r,pxx.petroR50Err_r,pxx.petroR90_r,pxx.petroR90Err_r,pz.z,pz.zErr,pz.nnCount,pz.photoErrorClass 
FROM GalaxyTag as px, 
     Photoz as pz, 
     PhotoObjAll as pxx     
WHERE px.modelMag_r<=17.77 
AND px.type=3 
AND px.mode=1
AND ((px.flags_r & 0x10000000)!=0)
AND ((px.flags_r & 0x8100000c00a0)=0)
AND (((px.flags_r & 0x400000000000)=0) 
OR (px.psfmagerr_r<=0.2))
AND (((px.flags_r & 0x100000000000)=0) 
OR (px.flags_r & 0x1000)=0)
AND px.ObjID=pz.ObjID
AND pxx.ObjID=px.ObjID
AND px.specObjID=0
\end{lstlisting}
The methods used to calculate the photometric redshift estimates retrieved with this query are described in \cite{Beck+16}.
\item From the sample of galaxies without redshifts described in the previous item, we selected those galaxies that satisfy $r>0$, $g-r<3$. Magnitudes were corrected for extinction and transformed to the AB system \citep{SDSStoAB}. Some of these galaxies were found to have their redshifts available in the NED. In Table~\ref{tab:newz} we quote the redshifts found for the photometric galaxies around CG candidates. We corrected the redshifts to the CMB rest frame using the standard equation
\[z_{\rm CMB}=z_{NED}+z_{a}[\sin{b} \sin{b_{a}}+ \cos{b}  \cos{b_{a}} \cos{(l_{a}-l)} ],\]
where $l$ and $b$ are the galactic coordinates, $l_{a}=264.14^{\circ}$, $b_{a}=48.26^{\circ}$, and $z_{a}=371.0/c$ (subscript $a$ is the abbreviation of $apex$) and $c$ is the velocity of light in  vacuum in $\rm km/s$.

\end{itemize}

\begin{table}
\caption{List of galaxies in the \citealt{tempel17} catalogue around potential compact groups that were visually classified as Part of a Galaxy (PofG). 
\label{tab:pofg}}
\begin{center}
\begin{tabular}{rc}
\hline
\hline
&SDSSObjID \\
\hline
\hline
1& 1237651823246573749\\
2& 1237661873468080155\\
3& 1237654879128060006\\
4& 1237651822173552811\\
5& 1237651252561379426\\
6& 1237657857676607536\\
7& 1237661816558714927\\
8& 1237648704047284236\\
9& 1237665330921013326\\
10& 1237661069261734123\\
11& 1237648722301419657\\
12& 1237654601570451465\\
13& 1237654601570451467\\
14& 1237658491206041662\\
15& 1237667915952816188\\
16& 1237658424093769746\\
17& 1237668273507729574\\
18& 1237665025444282416\\
19& 1237651251478724811\\
20& 1237665441511964723\\
21& 1237671991339516047\\
\hline
\end{tabular}
\end{center}
\end{table}

\begin{table}
\caption{Redshift corrected to the CMB rest frame for $63$ photometric galaxies in the SDSS DR14 extracted from the NASA/IPAC Extragalactic Database (NED). \label{tab:newz}}
\begin{center}
\small
\begin{tabular}{rccc}
\hline
\hline
& ObjID & $z_{\rm NED}$ & $z_{\rm CMB}$\\
\hline
\hline
1 & 1237665566604787742 & 0.061476 & 0.062265 \\
2 & 1237658609296146595 & 0.056870 & 0.057201 \\
3 & 1237664130484273189 & 0.142020 & 0.143180 \\
4 & 1237664130484273365 & 0.065240 & 0.066400 \\
5 & 1237664130484207707 & 0.068550 & 0.069710 \\
6 & 1237664130484207928 & 0.069491 & 0.070651 \\
7 & 1237661850938572946 & 0.087930 & 0.088548 \\
8 & 1237657608567783547 & 0.057460 & 0.058040 \\
9 & 1237657595683078689 & 0.058420 & 0.058998 \\
10 & 1237657608567783553 & 0.057460 & 0.058041 \\
11 & 1237651272419508510 & 0.048090 & 0.048565 \\
12 & 1237661850928676959 & 0.024390 & 0.025164 \\
13 & 1237661872399646878 & 0.094300 & 0.095075 \\
14 & 1237661872399646879 & 0.094300 & 0.095076 \\
15 & 1237655373572866101 & 0.028140 & 0.028216 \\
16 & 1237661872949297333 & 0.092020 & 0.092573 \\
17 & 1237661872949297281 & 0.090300 & 0.090854 \\
18 & 1237667734492020911 & 0.014460 & 0.015493 \\
19 & 1237665531723317688 & 0.037763 & 0.038075 \\
20 & 1237668670253039811 & 0.071470 & 0.071932 \\
21 & 1237668294985842855 & 0.162200 & 0.163290 \\
22 & 1237651753457483933 & 0.071500 & 0.072612 \\
23 & 1237658492269494418 & 0.047310 & 0.048481 \\
24 & 1237653617470603328 & 0.045010 & 0.045497 \\
25 & 1237655108374757517 & 0.049380 & 0.049877 \\
26 & 1237665443126116438 & 0.051652 & 0.052523 \\
27 & 1237661416068546653 & 0.036950 & 0.037288 \\
28 & 1237657856064618588 & 0.024980 & 0.025678 \\
29 & 1237648720164880479 & 0.082030 & 0.083096 \\
30 & 1237648720164880703 & 0.084400 & 0.085466 \\
31 & 1237654605873348678 & 0.075669 & 0.076827 \\
32 & 1237651822716715274 & 0.119100 & 0.119615 \\
33 & 1237667323797766247 & 0.026822 & 0.027721 \\
34 & 1237667323797766298 & 0.072250 & 0.073149 \\
35 & 1237667323797766905 & 0.022800 & 0.023697 \\
36 & 1237661387063885956 & 0.080280 & 0.080741 \\
37 & 1237658311867105407 & 0.059360 & 0.059821 \\
38 & 1237661465452740882 & 0.048000 & 0.048192 \\
39 & 1237659329781367066 & 0.066270 & 0.066296 \\
40 & 1237664871900643452 & 0.050148 & 0.051053 \\
41 & 1237657630057562291 & 0.061040 & 0.061713 \\
42 & 1237654880205209753 & 0.091600 & 0.092229 \\
43 & 1237662199355146310 & 0.051142 & 0.051814 \\
44 & 1237667550347460631 & 0.025124 & 0.026180 \\
45 & 1237667550347460635 & 0.025124 & 0.026180 \\
46 & 1237662637443842271 & 0.045111 & 0.045687 \\
47 & 1237651271888208191 & 0.029890 & 0.030349 \\
48 & 1237654604253233186 & 0.041870 & 0.043081 \\
49 & 1237665351854915780 & 0.069428 & 0.070044 \\
50 & 1237654601018507679 & 0.028610 & 0.029593 \\
51 & 1237654604776407247 & 0.027900 & 0.028882 \\
52 & 1237655691404968131 & 0.039110 & 0.039771 \\
53 & 1237655691404902688 & 0.072650 & 0.073313 \\
54 & 1237648722841043108 & 0.078600 & 0.079802 \\
55 & 1237667253462892735 & 0.025657 & 0.026634 \\
56 & 1237651822717239556 & 0.043920 & 0.044412 \\
57 & 1237657630599872534 & 0.027052 & 0.027669 \\
58 & 1237674650998145068 & 0.024140 & 0.025366 \\
59 & 1237667254015688747 & 0.012929 & 0.013876 \\
60 & 1237671751364313102 & 0.008382 & 0.008707 \\
61 & 1237662335183225035 & 0.134000 & 0.134208 \\
62 & 1237665548887261200 & 0.029552 & 0.030330 \\
63 & 1237667541753266247 & 0.024240 & 0.025276 \\
\hline
\end{tabular}
\end{center}
\end{table}

 \section{Non-contaminating galaxies}
\label{nope}
\begin{figure}
\begin{center}
\includegraphics[width=\hsize,trim=20 100 20 80,clip]{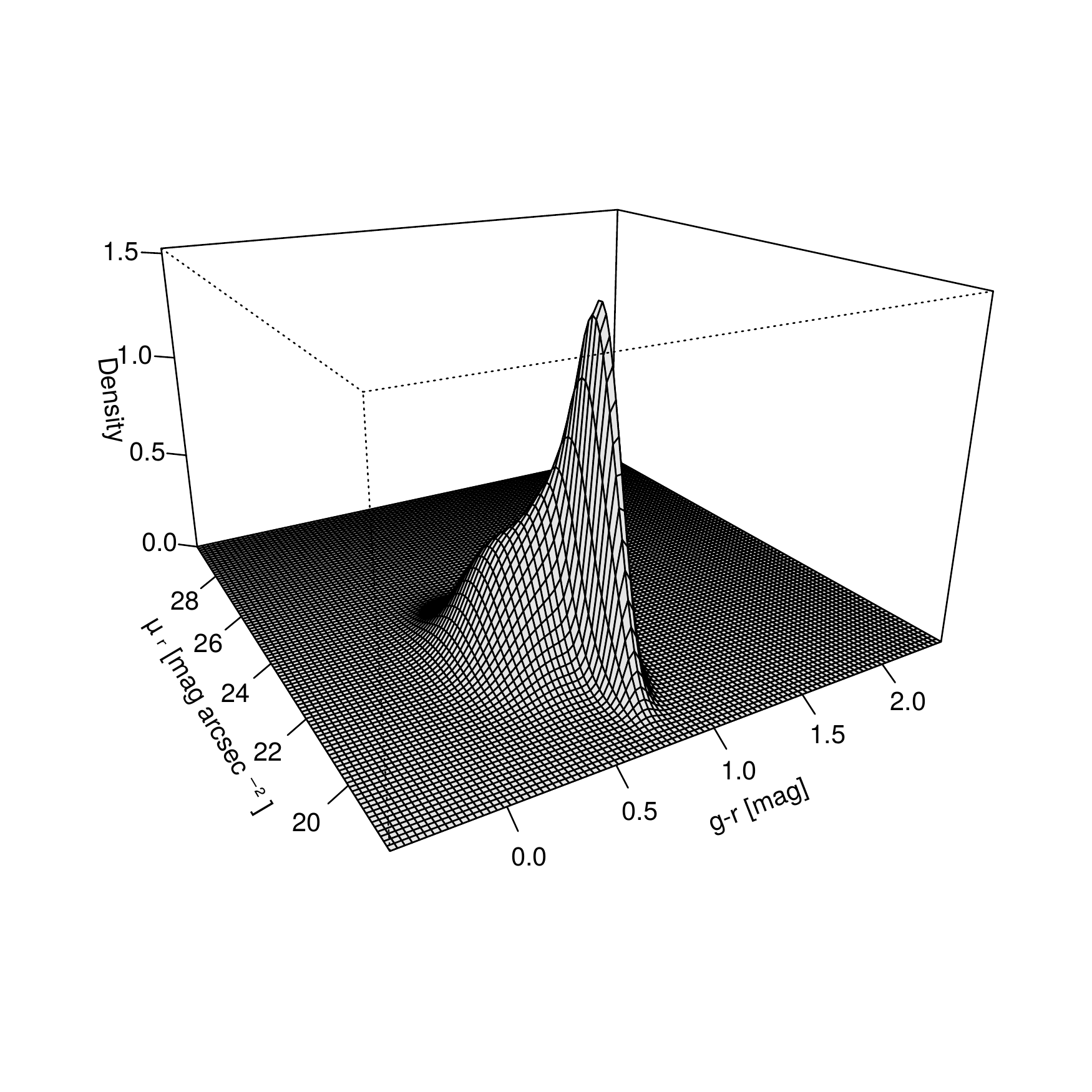}
\includegraphics[width=\hsize]{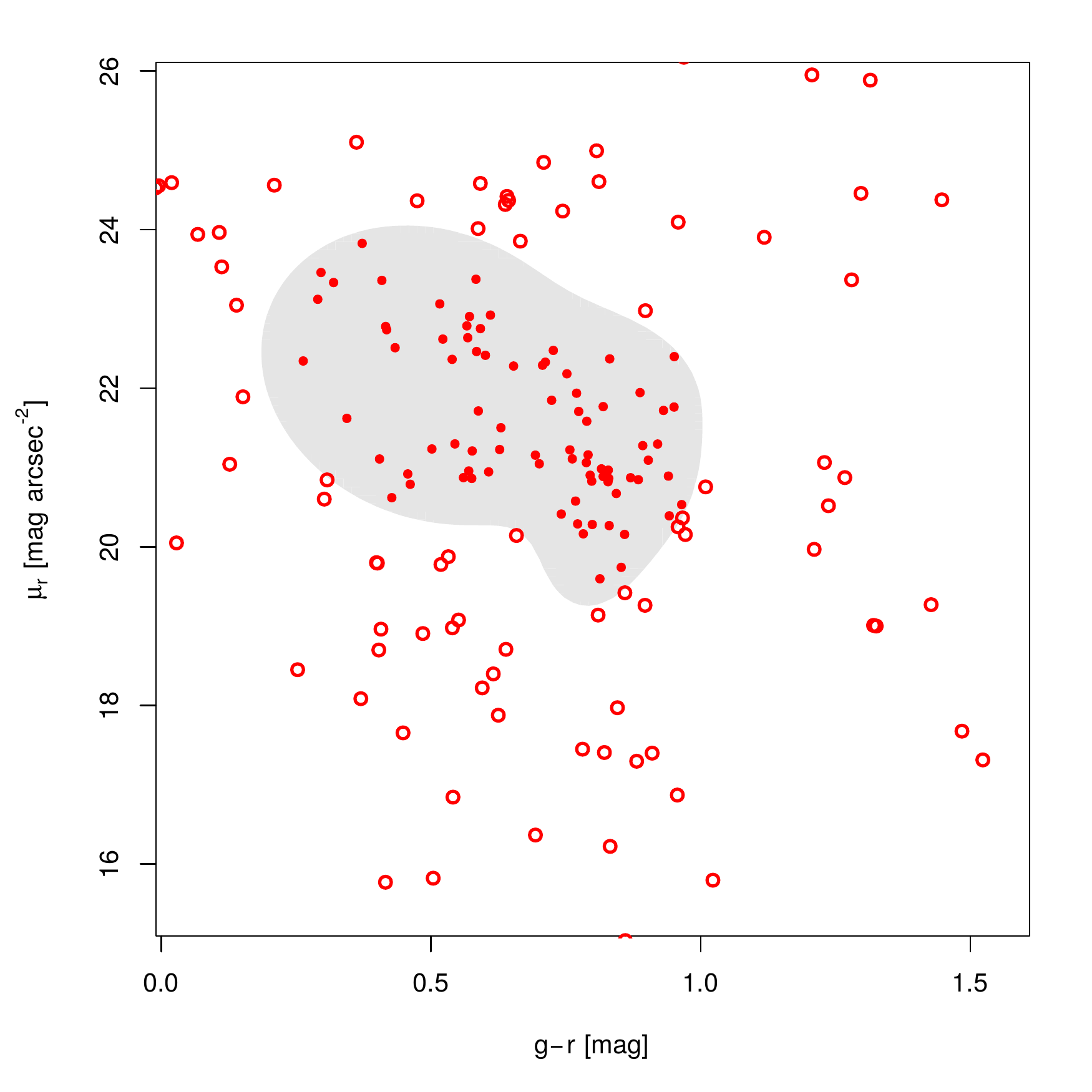}
\caption{\label{fc1} Upper plot: Perspective plot of the bivariate density estimate in the surface brightness vs colour plane for the spectroscopic galaxies in the SDSS DR14 that are in the same redshift and magnitude ranges as those \texttt{HMCG}s that contain potentially contaminating photometric galaxies.  
Lower plot: Grey area encloses  $95\%$ of the spectroscopic galaxies.
Red filled dots are photometric galaxies in the same magnitude range of their host CGs, and that are considered potential sources of contamination (inside the grey region), while red open dots are galaxies classified as non-contaminating objects (outside the grey region)(see text for details).
} 
\end{center}
\end{figure}


In this appendix, we focus on the $185$ photometric galaxies considered as potential sources of contamination of \texttt{HMCG}s in~Sect.~\ref{sec:HMCG}. We compared the photometric properties of these galaxies with those of the spectroscopic galaxies around the same  \texttt{HMCG}s that they might be contaminating. 

For each group with a potential contaminating galaxy, we selected spectroscopic galaxies from the SDSS DR14 that are
\begin{itemize}
\item inside a circle of $5$ degrees radius from the \texttt{HMCG} centre,
\item within a $3$ magnitude range from the  \texttt{HMCG} first-ranked galaxy,
\item and within $1000 \rm \, km s^{-1}$ of the \texttt{HMCG} median radial velocity.
\end{itemize}
We measured the surface brightnesses ($\mu_r$) and observer-frame colours ($g-r$) of each spectroscopic galaxy and of the potentially contaminating galaxies.  The surface brightness of each galaxy was computed using the Petrosian radius provided by the SDSS pipeline. After analysing the 
surface brightness~$-$~colour relation, we used the $Mclust$ package \citep{mclust} of R software to perform a density estimation based on finite Gaussian mixture modelling. From its application to our data in the surface brightness~$-$~colour plane, we observed that, according to the Bayesian Information Criterion, a proper density representation can be obtained using a mixture of four Gaussian functions for a model with ellipsoidal distribution that allows variable volume, shape, and orientation. The best fitted parameters for the obtained Gaussian mixture modelling are quoted in Table~\ref{tab:gauss}, while the perspective plot of the resulting surface density is shown in the upper panel of Fig.~\ref{fc1}.

Using this surface density, we defined the isodensity contour that enclosed  $95\%$ of the spectroscopic galaxies in the surface brightness~$-$~colour plane. 
This region is represented as the grey shaded area in the lower panel of Fig.~\ref{fc1}. We also show the scatter plot for the potential sources of contamination (red dots). Photometric galaxies lying outside the grey region are considered non-contaminating galaxies (empty red dots). This criterion allowed us to discard $104$ galaxies.

Finally, as described in Sect.~\ref{sec:HMCG}, after a visual inspection of the remaining $81$ objects, we confirm that $13$ were actually objects misclassified as galaxies by the SDSS pipeline. Therefore, only $68$ galaxies are kept as potential sources of contamination. 
The SDSS object identification numbers of these $68$ galaxies are quoted in Table~\ref{tab:intruder}. 

\begin{table}
\caption{$Mclust$ best fitted parameters obtained for the bivariate density estimate from four Gaussian functions in the surface brightness~$-$~colour plane.
The MP values quoted in the second column are the mixing proportions for each component of the mixture model. For the mean vectors and covariance matrices, the $x$ variable corresponds to the galaxy colour $g-r$ , while the $y$ variable is the surface brightness $\mu_r$. \label{tab:gauss}}
\begin{center}
\setlength{\tabcolsep}{2pt}
\begin{tabular}{cccccccc}
\hline
\hline
Gaussian & MP & \multicolumn{2}{c}{Mean vector} & & \multicolumn{3}{c}{Covariance matrix}\\
\cline{3-4}
\cline{6-8}
& &\texttt{$\bar{x}$} & \texttt{$\bar{y}$} & & \texttt{$\sigma_{xx}$}& \texttt{$\sigma_{yy}$}& \texttt{$\sigma_{xy}=\sigma_{yx}$} \\
\hline
1 & 0.220 & 0.745 & 21.776 & & 0.0132 & 0.2884 & ~0.0046 \\
2 & 0.451 & 0.500 & 22.179 & & 0.0239 & 0.8453 & -0.0209 \\
3 & 0.262 & 0.827 & 21.024 & & 0.0044 & 0.2565 & ~0.0011 \\
4 & 0.067 & 0.812 & 20.216 & & 0.0024 & 0.2212 & ~0.0051 \\
\hline
\hline
\end{tabular}
\end{center}
\end{table}

 \begin{table}
\caption{List of galaxies in the SDSS DR14 photometric catalogue around compact groups that are potential sources of contamination
\label{tab:intruder}}
\begin{center}
\begin{tabular}{rcrc}
\hline
\hline
&SDSSObjID  & & SDSSObjID \\
\hline
\hline
    1 &   1237651272419509016 & 35 &   1237661850928611367  \\ 
    2 &   1237660635999764658 & 36 &   1237665442601107677  \\ 
    3 &   1237661971727515740 & 37 &   1237665442601107712  \\ 
    4 &   1237661416068546838 & 38 &   1237665442601107767  \\ 
    5 &   1237662199356850525 & 39 &   1237665548349931560  \\ 
    6 &   1237665442601107571 & 40 &   1237662336256442628  \\ 
    7 &   1237667781744328983 & 41 &   1237662337865482596  \\ 
    8 &   1237651252027326626 & 42 &   1237662337865482480  \\ 
    9 &   1237653617470603455 & 43 &   1237662636911034773  \\ 
   10 &   1237651822716649609 & 44 &   1237665179521319111  \\ 
   11 &   1237651753457418421 & 45 &   1237665533325541581  \\ 
   12 &   1237654948378771564 & 46 &   1237665533325541619  \\ 
   13 &   1237657857144717348 & 47 &   1237665429168783623  \\ 
   14 &   1237660412650193181 & 48 &   1237665179522760752  \\ 
   15 &   1237654605873348872 & 49 &   1237661949196763146  \\ 
   16 &   1237658493347758315 & 50 &   1237664295293812934  \\ 
   17 &   1237655693557891393 & 51 &   1237667538550456389  \\ 
   18 &   1237659162276135058 & 52 &   1237668273507729489  \\ 
   19 &   1237658492280242289 & 53 &   1237668272440672468  \\ 
   20 &   1237661971727515766 & 54 &   1237668495776219326  \\ 
   21 &   1237660671429967980 & 55 &   1237667448873549879  \\ 
   22 &   1237660671429967981 & 56 &   1237667782270648349  \\ 
   23 &   1237661387617730866 & 57 &   1237667911138279447  \\ 
   24 &   1237662199355146533 & 58 &   1237668298738172072  \\ 
   25 &   1237662528995852513 & 59 &   1237668289620672636  \\ 
   26 &   1237661086961107151 & 60 &   1237668611197567138  \\ 
   27 &   1237662636370100371 & 61 &   1237668336862691636  \\ 
   28 &   1237665429170618580 & 62 &   1237668336862691739  \\ 
   29 &   1237661418747134094 & 63 &   1237671991336239245  \\ 
   30 &   1237665548351438938 & 64 &   1237654390570680607  \\ 
   31 &   1237665442601042133 & 65 &   1237658205048930492  \\ 
   32 &   1237662194533531763 & 66 &   1237665548351438942  \\ 
   33 &   1237665533338255410 & 67 &   1237665567153127563  \\ 
   34 &   1237662337865417030 & 68 &   1237668272440672437  \\
\hline
\end{tabular}
\end{center}
\end{table}

\section{New CG catalogue in the SDSS}
\label{catalogue}
In this appendix we present the tables of \texttt{HMCG}s identified in the SDSS. In Table~\ref{tab:groups} we show the content for the table containing $462$ \texttt{HMCG}s, while Table~\ref{tab:members} shows the content for the table that contains the information for the $2070$ galaxy members.

\begin{table*}
\caption{Compact groups identified in the SDSS DR12 using the \texttt{modified} algorithm.\label{tab:groups}}
\begin{center}
\begin{tabular}{cccccccccc}
\hline
\hline
\texttt{HMCG}id & $N$ & RA & Dec & Redshift  & $\Theta_{\rm G}$ & $\mu_r$ & $\sigma_v$ & $r_{\rm b}$ & Flag\\
 &  & [deg] & [deg] &  &  $ [arcmin]$ & [$mag \, arcsec^{-2}$] & [$km \, s^{-1}$] & & \\
\hline
\hline
 1 &        4 &     114.842 &      45.103 &  0.078292 &     2.620 &    25.018 &     348.679 &    14.719 &        0 \\
 2 &        4 &     116.577 &      22.020 &  0.046817 &     4.090 &    25.289 &     274.092 &    14.289 &        0 \\
 3 &        4 &     117.288 &      21.762 &  0.024476 &     9.769 &    26.134 &     136.700 &    13.359 &        0 \\
 4 &        5 &     117.794 &      50.217 &  0.021810 &     5.095 &    23.789 &     409.803 &    12.839 &        0 \\
 5 &        4 &     118.709 &      45.713 &  0.053666 &     1.065 &    22.685 &     515.812 &    14.645 &        0 \\
\vdots & \vdots & \vdots & \vdots & \vdots &  \vdots & \vdots & \vdots & \vdots & \vdots\\
458 &        7 &     243.981 &      38.537 &  0.034782 &     6.091 &    26.202 &     138.779 &    14.577 &        1 \\
459 &        4 &     244.410 &      50.641 &  0.041362 &     4.748 &    25.610 &     552.432 &    14.146 &        1 \\
460 &        5 &     245.283 &      13.159 &  0.034352 &     3.715 &    24.912 &     247.585 &    14.328 &        1 \\
461 &        4 &     247.547 &      36.247 &  0.075265 &     3.671 &    25.664 &     383.398 &    14.479 &        1 \\
462 &        5 &     250.332 &      13.424 &  0.050796 &     4.130 &    24.858 &     268.188 &    13.764 &        1 \\
\hline
\multicolumn{10}{p{.85\textwidth}}{Notes. \texttt{HMCG}id: group ID, $N$: number of galaxy members, RA: group centre right ascension (J2000), Dec: group centre declination (J2000), Redshift: group CMB redshift, $\Theta_{\rm G}$: angular diameter of the smallest circumscribed circle, $\mu_r$: r-band group surface brightness, $\sigma_v$: radial velocity dispersion, $r_{\rm b}$: r-band observer-frame model apparent magnitude of the group brightest galaxy, Flag: $0=$ clean groups, $1=$ potentially contaminated groups. Groups in this table were identified using the \texttt{modified} algorithm presented in this work. 
This table  is available in electronic form.
}
\end{tabular}
\end{center}
\end{table*}

\begin{table*}
\caption{Galaxy members of compact groups identified in SDSS DR12 using the \texttt{modified} algorithm.\label{tab:members}}
\begin{center}
\begin{tabular}{ccccccc}
\hline
\hline
\texttt{HMCG}id & $N_m$ & RA & Dec & Redshift & $r$ & $g$\\
 &  & [deg] & [deg] &  &  & \\
\hline
\hline
   1 &        1 &     114.844 &      45.118 &  0.079785 &    14.719 &    15.661 \\
   1 &        2 &     114.840 &      45.124 &  0.077575 &    16.380 &    16.885 \\
   1 &        3 &     114.850 &      45.082 &  0.078914 &    16.813 &    17.693 \\
   1 &        4 &     114.819 &      45.088 &  0.077670 &    17.100 &    17.960 \\
   2 &        1 &     116.553 &      21.996 &  0.045715 &    14.289 &    15.159 \\
   2 &        2 &     116.610 &      22.034 &  0.047243 &    14.982 &    15.809 \\
   2 &        3 &     116.543 &      22.006 &  0.047494 &    15.822 &    16.678 \\
   2 &        4 &     116.576 &      22.002 &  0.046392 &    16.672 &    17.471 \\
   3 &        1 &     117.331 &      21.833 &  0.024654 &    13.359 &    14.210 \\
   3 &        2 &     117.222 &      21.747 &  0.023674 &    13.511 &    14.316 \\
   3 &        3 &     117.207 &      21.733 &  0.024548 &    15.695 &    16.442 \\
   3 &        4 &     117.375 &      21.753 &  0.024403 &    15.774 &    16.067 \\
\vdots & \vdots & \vdots & \vdots & \vdots & \vdots & \vdots \\
\hline
\multicolumn{7}{p{.55\textwidth}}{Notes: \texttt{HMCG}id: group ID, $N_m$: galaxy index, RA: right ascension (J2000), Dec: declination (J2000), z: CMB redshift, $r$, $g$: r-band and g-band observer-frame model apparent magnitudes corrected for extinction in the AB system. Galaxies of each group are ordered by their apparent magnitudes from brightest to faintest. 
This table is available in electronic form.
}
\end{tabular}
\end{center}
\end{table*}

\label{lastpage}
\end{appendix}
\end{document}